\documentclass{aa}
\usepackage[varg]{txfonts}
\usepackage{newtxtext,newtxmath}	
\usepackage[T1]{fontenc}
\usepackage{ae,aecompl}
\usepackage{amsmath}
\usepackage{amsfonts}
\usepackage{amssymb}
\usepackage{graphicx}
\usepackage{hyperref}
\hypersetup{colorlinks=true,linkcolor=blue,citecolor=blue,filecolor=blue,urlcolor=blue}

\def\aap{A$\&$A}

\def\apjl{ApJ Letters}
\def\apjs{ApJS}

\newcommand{\me}{\, {\rm M}_{\oplus}}
\newcommand{\msun}{\, {\rm M}_{\odot}}
\newcommand{\au}{\, {\rm au}}
\newcommand{\trap}{{Trappist-1}}
\newcommand{\mpl}{\, {m}_{\rm p}}

\begin{document}
\title{Pebbles versus Planetesimals: The case of \trap}
\author{G. A. L. Coleman\thanks{Email: gavin.coleman@space.unibe.ch}, A. Leleu\thanks{CHEOPS Fellow}, Y. Alibert, W. Benz\\}
\institute{Physikalisches Institut, Universit\"at Bern, Gesellschaftsstr.\ 6, 3012 Bern, Switzerland}

\titlerunning{Pebbles vs Planetesimals}
\authorrunning{Coleman et al}
\date{}
\abstract{We present a study into the formation of planetary systems around low mass stars similar to \trap, through the accretion of either planetesimals or pebbles.
The aim is to determine if the currently observed systems around low mass stars could favour one scenario over the other. 
To determine these differences, we ran numerous N-body simulations, coupled to a thermally evolving viscous 1D disc model, and including prescriptions for planet migration, photoevaporation, and pebble and planetesimal dynamics.
We mainly examine the differences between the pebble and planetesimal accretion scenarios, but we also look at the influences of disc mass, size of planetesimals, and the percentage of solids locked up within pebbles.
When comparing the resulting planetary systems to \trap, we find that a wide range of initial conditions for both the pebble and planetesimal accretion scenarios can form planetary systems similar to \trap, in terms of planet mass, periods, and resonant configurations.
Typically these planets formed exterior to the water iceline and migrated in resonant convoys into the inner region close to the central star.

When comparing the planetary systems formed through pebble accretion to those formed through planetesimal accretion, we find a large number of similarities, including average planet masses, eccentricities, inclinations and period ratios.
One major difference between the two scenarios was that of the water content of the planets.
When including the effects of ablation and full recycling of the planets' envelope with the disc, the planets formed through pebble accretion were extremely dry, whilst those formed through planetesimal accretion were extremely wet.
If the water content is not fully recycled and instead falls to the planets' core, or if ablation of the water is neglected, then the planets formed through pebble accretion are extremely wet, similar to those formed through planetesimal accretion.
Should the water content of the \trap ~planets be determined accurately, this could point to a preferred formation pathway for planetary systems, or to specific physics that may be at play.

}
\maketitle

\section{Introduction}
\label{sec:intro}
The recent discovery of seven Earth-sized planets orbiting the low mass star \trap ~\citep{GillonTrappist17,LugerTrappist1-h} has led to many questions about the formation and evolution of such a complex system.
Not only are all seven planets orbiting close to their parent star (orbital periods $\le$ 20 d), but they also all appear to form a resonant chain, such that their orbital periods are near integer ratios of each other.
The formation of such resonant chains are a natural outcome of interactions between the planets and their nascent protoplanetary discs \citep{cressnels}.
These resonant chains have been observed in other compact planetary systems \citep{Kepler11,Fabrycky2014,Mills16}, and have also been formed in complex planet formation simulations involving multiple bodies \citep{Hellary,ColemanNelson14,ColemanNelson16,ColemanNelson16b}.
Whilst \trap ~may be the most high-profile planetary system around low mass stars, it is interesting to note that a number of similar planetary systems have also been recently observed.
For example, two planets with periods less than 5 days have been confirmed around YZ Ceti, with a third planet still awaiting confirmation \citep{Astudillo-Defru17,Robertson18}.
GJ 1132 \citep{Berta-Thompson15,Bonfils18_GJ1132} and GJ 3323 \citep{Astudillo-Defru17_GJ3323} each have two super-Earths orbiting close to their central star, whilst Proxima \citep{Anglada2016} and Ross 128 \citep{Bonfils18_Ross128} each contain a planet that is very similar in mass and period to \trap ~g.
More recently, planets have also been observed around Barnard's Star \citep{Ribas18}, LHS 1140 \citep{Dittmann17,Ment19}, and GJ 1214 \citep{Luque18}, significantly increasing the number of planets observed around low mass stars.

Following the discovery of Proxima b \citep{Anglada2016}, \citet{ColemanProxima17} presented numerous formation scenarios for such a planet orbiting such a low-mass star.
These scenarios ranged from {\it in situ} formation, to migration of a single or multiple planetary embryos from outside the iceline after accreting either planetesimals or pebbles.
They showed that each scenario yielded subtly different observational signatures such as multiplicity, planet composition and orbital architectures.
{\it In situ} formation produced numerous volatile-poor Earth-sized planets with little evidence for resonant chains; migration of a single embryo from outside the iceline formed a single volatile-rich Earth-sized planet on a circular orbit; whilst the migration of multiple embryos formed numerous Earth-sized planets rich in volatiles, often displaying mean-motion resonances (MMRs) between neighbouring planets.
More recently \cite{AlibertBenz17} studied the formation and composition of planets around low mass stars, finding that close-in planets have similar masses and radii (peaking at $\sim 1 R_{\oplus}$), and also that the properties of the protoplanetary disc and their correlation with the stellar mass are important in determining the characteristics of the planet, e.g. water content.
Since these papers were either only aiming to form a single planet, or only involved single-planet-in-a-system simulations, they did not address the formation of such a complex system as \trap.

A scenario that has recently been proposed for the formation of the \trap ~planetary system through pebble accretion is outlined in \cite{Ormel17Trappist}.
In their scenario, they assume that the planetary embryos form at the water iceline in the disc, after millimeter- or centimeter-sized particles (pebbles) have accumulated there.
Once a planetary embryo forms, it accretes the surrounding pebbles before migrating in towards the central star.
As the embryo migrates inwards,  it accretes dry pebbles, further increasing its mass, whilst a new embryo forms at the iceline and goes through the same process.
The planets then migrate to the inner edge of the disc, where their migration ceases, allowing the planets to enter into first-order mean motion resonances.
As the disc disperses, the first order resonances can be broken allowing the planets to dynamically rearrange into new configurations \citep{ColemanNelson16,Izidoro17}.
Recently \citet{Schoonenberg19} explored this scenario in a more quantitative setting.
They found that this method was unable to form planetary systems similar to \trap.
However, instead of forming planetary embryos one after another, they were able to form planetary systems similar to \trap~ if multiple planetary embryos formed at the iceline on short time-scales (e.g. 1000 y).
These embryos could then mutually interact and accrete dry and wet pebbles whilst slowly migrating closer to the central star.
The planets in the simulated systems then compared favourably to \trap~ in terms of planet masses and water fractions.

In this paper we use up-to-date models of planet formation utilising either planetesimal or pebble accretion, in studying the formation of planetary systems around low mass stars, with a specific goal in forming planetary systems similar to \trap.
We use the Mercury-6 symplectic integrator to compute the dynamical evolution and collisional accretion of planetary embryos and planetesimals \citep{Chambers}.
This is combined with a 1-D viscous disc model that incorporates thermal evolution through stellar irradiation, viscous heating and blackbody cooling.
The simulations also incorporate up-to-date prescriptions for planet migration, enhanced planetesimal capture by planetary atmospheres, and gas disc dispersal through photoevaporation on Myr time-scales.
For the planetesimal accretion scenario we embed the initial planetary embryos amongst thousands of planetesimals, whilst for the pebble accretion scenario we include the pebble accretion and evolution models of \cite{Lambrechts14}.
Interestingly, we find that both scenarios are able to form planetary systems comparable to \trap ~quite well, with very little to separate the two models in terms of observational features.
There do exist some differences between the planetary systems formed in the two scenarios, that could with future observations of \trap ~and other planetary systems around low mass stars, inform as to the preferred mode of planet formation for these systems.

This paper is organised as follows.
We briefly describe the physical model in Sect. \ref{sec:models}.
We then examine the formation of planetary systems through the accretion of planetesimals in Sect. \ref{sec:planetesimals}.
In Sect. \ref{sec:pebbles} we use pebble accretion instead of planetesimal accretion to form the planetary systems.
We then compare the results of the planetesimal and pebble accretion scenarios in Sect. \ref{sec:comparisions}, and then we draw our conclusions in Sect. \ref{sec:conclusions}.

\section{Physical Model}
\label{sec:models}
The physical model we adopt for this study is based on the planet formation models of \citet{ColemanNelson14,ColemanNelson16}.
These models run N-body simulations using the Mercury-6 symplectic integrator \citep{Chambers}, adapted to include the disc models and physical processes described below.

\noindent
(i) We solve the standard diffusion equation for a 1D viscous $\alpha$-disc model \citep{Shak,Lynden-BellPringle1974}. 
Disc temperatures are calculated by balancing black-body cooling against viscous heating and stellar irradiation. The viscous parameter $\alpha_{\rm visc}=1 \times 10^{-3}$ throughout most of the disc, but increases to $\alpha_{\rm active}=0.005$ in regions where $T \ge 1000$ K to mimic the fact that fully developed turbulence can develop in regions where the temperature exceeds this value \citep{UmebayashiNakano1988,DeschTurner2015}.
 
\noindent
(ii) The final stages of disc removal occur through a photoevaporative wind.
We use a standard photoevaporation model for most of the disc evolution \citep{Dullemond}, corresponding to a photoevaporative wind being launched from the upper and lower disc surfaces. Direct photoevaporation of the disc is switched on during the final evolution phases when an inner cavity forms in the disc, corresponding to the outer edge of the disc cavity being exposed directly to the stellar radiation \citep{Alexander09}.
 
\noindent
(iii) The N-body simulations consist of a number of planetary embryos that can mutually interact gravitationally and collide.
In addition, some models also include planetesimals (bodies with radii either $100$ m $ \le R_{\rm p} \le 1$ km). Planetesimals orbiting in the gaseous protoplanetary disc experience size dependent aerodynamic drag \citep{Adachi,Weidenschilling_77}.
Collisions between protoplanets and other protoplanets or planetesimals always result in perfect merging.
Planetesimal-planetesimal interactions and collisions are neglected for reasons of computational speed.
 
\noindent
(iv) We use the torque formulae from \citet{pdk10,pdk11} to simulate type I migration due to Lindblad and corotation torques acting on the planetary embryos.
Corotation torques arise from both entropy and vortensity gradients in the disc, and the possible saturation of these torques is included in the simulations.
The influence of eccentricity and inclination on the migration torques, and of eccentricity and inclination damping are included \citep{Fendyke,cressnels}.

\noindent
(v) The effective capture radius of planetary embryos accreting planetesimals is enhanced by atmospheric drag as described in section 2.5 of \citet{Inaba}.
This prescription from \citet{Inaba} provides an estimate of the atmosphere density as a function of radius, $\rho(R)$, which is then used to calculate a critical radius of the planet, that if a planetesimal crosses, it will be accreted.

The inner boundary of the computational domain is located just inside $0.01 \au$.
Any planets whose semi-major axes are smaller than this boundary radius are removed from the simulation and are assumed to have hit the star.

We neglect the accretion of gaseous envelopes in these calculations, as preliminary tests using the self-consistent 1D gas accretion model outlined in \cite{CPN17} showed that planets with relevant masses to these simulations, could only accrete $\sim 1 \%$ of their total mass in gas, over the course of the disc lifetime.
These values were found during both the solid accretion phase (where solid accretion luminosity would account for the planet's luminosity), and the post solid accretion phase, where gas would accrete onto the cores much slower than it would take for the planets to migrate into the inner hotter regions of the disc where the accretion of gas is inhibited.
Since the mass of the envelope is negligible compared to the total mass of the planet, it would have little effect on the N-body interactions between planets.
Also, once the disc had fully dispersed, such small gas envelope masses would likely be stripped away on short time-scales by the irradiation from the central star \citep{Owen17,Jin18}.
Given these lack of effects of the gaseous envelope, we neglect their accretion so as to save on computation time.

For the disc model, we use one similar to that constructed for the study of the formation of Proxima b \citep{ColemanProxima17}, since the stellar properties of Proxima Centauri and \trap ~are similar, i.e. both low mass M-dwarfs with similar masses, temperatures and radii.
By adopting a similar approach to the construction of the minimum mass solar nebular \citep{Hayashi}, \citet{ColemanProxima17} constructed a disc appropriate for Proxima Centauri that contained $\sim 4.5\%$ of Proxima's mass.
Following \citet{WilliamsCieza11} who quote a relationship between disc mass and radius, and by comparing this with the mass and radius of the Solar System, they find a radius for the fiducial disc around Proxima to have a radius of 10 $\au$.
We use similar values for our disc models presented here with Table \ref{tab:discparameters} presenting the disc parameters used in the simulations.
We initialise the disc surface density and temperature profiles using, $\Sigma_{\rm g}(r) = \Sigma_{\rm g}(1\au)(r/\au)^{-\alpha}$ and T(r) = T(1$\au$)(r/$\au)^{-\beta}$ respectively, where the values for $\Sigma_{\rm g}(1\au)$, T(1$\au$), $\alpha$ and $\beta$ can be found in Table \ref{tab:discparameters}.
For an explanation of the evolution of the disc, as well as the resulting migration tendencies for the planets, see section 3 of \cite{ColemanProxima17}.

\begin{table}
\centering
\caption{Disc and stellar model parameters}
\begin{tabular}{lc}
\hline
Parameter & Value\\
\hline
Disc inner boundary & 0.015 $\au$\\
Disc outer boundary & 10 $\au$\\
$\Sigma_{\rm g}$(1 $\au$) & $369 ~{\rm g/cm}^{-2}$\\
Surface density exponent $\alpha$ & -0.5\\
T(1 $\au$) & 55 K\\
Temperature exponent $\beta$ & -0.5\\
$\alpha_{\rm visc}$ & $1 \times 10^{-3}$\\
$\alpha_{\rm active}$ & $5 \times 10^{-3}$\\
Stellar Mass & $0.1\rm M_{\bigodot}$\\
$\rm R_{\rm S}$ & $1 \rm R_{\bigodot}$\\
$\rm T_{\rm S}$ & 3000 K\\
\hline
\end{tabular}
\label{tab:discparameters}
\end{table}
\section{Planetesimals}
\label{sec:planetesimals}
Numerous works have studied the formation of compact planetary systems via planetesimal accretion \citep{TerquemPapaloizou2007,McNeilNelson2010,Hellary, Cossou14,ColemanNelson14,ColemanNelson16}.
However none of these examined the formation of compact systems around low-mass stars similar to \trap.
More recently, \citet{ColemanProxima17} examined the formation of Proxima b through a number of different scenarios.
Some of the scenarios explored did use planetesimal accretion to form the planet, but were only an exploration of scenarios and only attempted to match observations of a single planet.
Hence, no detailed analysis of the multi-planet systems that they formed was carried out.
In this section therefore, we expand on the planetesimal formation scenarios presented in \citet{ColemanProxima17}, to run numerous simulations and explore how changing parameters such as planetesimal size, disc mass, and lifetime of the disc, affects the planetary systems that form, and whether they can form systems that are similar to \trap.

\begin{figure*}
\centering
\includegraphics[scale=0.8]{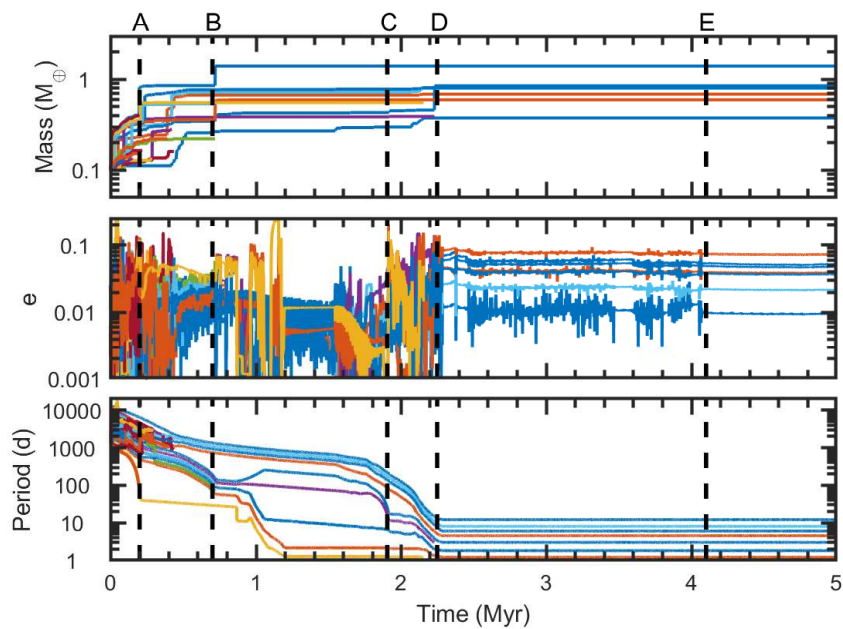}
\caption{Temporal evolution of planet masses (top), eccentricities (middle) and periods (bottom).
The noted letters indicate times of interesting events that are described in the text.}
\label{fig:pltml_sim}
\end{figure*}

\subsection{Initial Setup}
\label{sec:pltml_setup}
We use the disc model described above, and initially place 30 planetary embryos of mass 0.1$\me$ into the disc between 1 and 5$\au$.
The planetary embryos are embedded in amongst a swarm of 2000 planetesimals, that are spread between 0.5 and 5.5 $\au$, and have masses between 0.002 and 0.01 $\me$ depending on the total solid mass in the simulation.
We choose these values for the masses of planets/planetesimals and the numbers of each to allow the simulations to run on reasonable time-scales.
We assume that the mass locked within the planetary embryos and planetesimals accounts for half of the total solid mass of the disc, with the remaining 50$\%$ of the solid mass being set as small dust within the disc, that contributes to opacity calculations.
The effective physical radius of the planetesimals were set to either 100m or 1km.
We vary the initial solid surface density profile to examine the effect of changing the distributions of the solids, from being concentrated close to the inner regions of the disc and the iceline, to being moderately distributed throughout most of the disc.
To vary the distribution, we use a power-law surface density profile given by $\Sigma(R)\propto R^{-\alpha}$ where $\alpha$ takes the values, 1.5, 2, 2.5 or 3.
Initial eccentricities and inclinations are randomised according to a Rayleigh distribution, with scale parameters $e_0=0.01$ and $i_0=0.25^{\circ}$, respectively.
These eccentricities and inclinations will evolve over time due to interactions with planetary embryos, and through gas drag with the local gas disc.
We do not include the effects of stirring, either from turbulent gas, or mutual interactions with other planetesimals.
Since the interactions with the planetary embryos dominates the eccentricity and inclination evolutions, we do not expect the inclusion of such effects to have a significant effect.
We run all simulations for 10 Myr to allow the systems of formed planets to continue evolving through scattering and collisions after the dispersal of the protoplanetary discs.

For the disc parameters, we use disc masses of between 2.7 and 8$\%$ of the mass of \trap.
To account for different disc lifetimes, we modify the photoevaporation rate such that the $f41$ factor takes values of either, 0.0001, 0.001, 0.01 or 0.1.
This mainly affects the disc at the the end of its lifetime, as it determines when the disc begins to quickly disperse from the inside-out.
Finally, we run two versions of each set of parameters, where the planet positions and velocities are initialised using a different random number seed.

\subsection{Example Simulation}
\label{sec:pltml_example}
\begin{figure}
\centering
\includegraphics[scale=0.6]{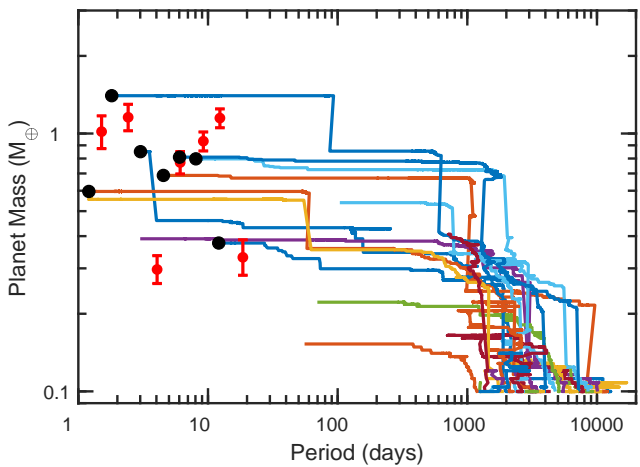}
\caption{Evolution of planet mass versus period for an example planetesimal accretion simulation.
Filled black circles represent final masses and periods for surviving planets.
The red dots indicates the periods and masses of the \trap ~planets with their appropriate error bars \citep{Grimm18}.}
\label{fig:pltml_sim_mvp}
\end{figure}

We now present an example simulation that produced a Trappist-like system.
Figure \ref{fig:pltml_sim} shows the temporal evolution of planet periods, masses and eccentricities of an example simulation that formed a Trappist-like system, and Fig. \ref{fig:pltml_sim_mvp} shows the systems' mass versus period evolution.
The black dots in Fig. \ref{fig:pltml_sim_mvp} represent the final planet masses and orbital periods, whilst the red points represent the \trap ~planets with appropriate errorbars.
The initial disc mass for this simulation was equal to $\sim4\%$ of the stellar mass, with $\sim11\me$ of solid material locked up in planetary embryos/planetesimals.

Over the first 0.1 Myr of the simulation, a number of planets accreted a significant number of planetesimals and raised their masses to 0.3$\me$.
These planets then began to undergo significant migration and started to migrate towards the central star. 
After another 0.1 Myr, two planets interacted with a swarm of planetesimals and quickly migrated into the inner regions of the disc where they became trapped at the outer edge of the active turbulent region, a region where fully developed turbulence increases the local $\alpha_{\rm visc}$ parameter \citep{UmebayashiNakano1988,DeschTurner2015}.
Once at the active turbulent region, these two planets collided giving rise to a 0.5$\me$ planet along with a swarm of planetesimals that had also become trapped due to the super-Keplerian velocity of the local gas, which reduced the gas drag acting on the planetesimals to zero.
The evolution of the two planets can be seen at the dashed line marked `A', and by the yellow and red lines in the bottom left of the bottom panel of Fig. \ref{fig:pltml_sim}.
After 0.3 Myr, a number of the outer planets converged at regions of zero migration, where their Lindblad torques balanced their corotation torques.
At these regions of zero migration, they cause planets to undergo relative convergent migration, leading to their capture into mean-motion resonance\footnote{Note here, we define that two planets are in resonance when their resonance angles are found to be consistently librating. This includes first and and second order resonances, and we neglect higher order resonances due to their scarcity.}, usually of first order.
Mutual interactions between planets also leads to a number of collisions, allowing planets to quickly grow to larger masses.

Typically as planets gain in mass, they begin to migrate inwards on shorter time-scales.
This is unless they are trapped at a region of zero migration \citep{Masset2006}.
At one of these regions, as a planet increases in mass, it's corotation torque will begin to saturate, allowing Lindblad torques to begin to dominate migration.
As the planet increases further in mass, its corotation torque will fully saturate, leaving only the Lindblad torques to drive migration, giving rise to short migration time-scales.
This scenario begins to happen for a number of more massive planets in the example simulation here, after $\sim0.4$ Myr.
As a planet of mass, $\mpl\sim 0.9\me$ accretes more planetesimals, it begins to saturate its corotation torque.
This allows the planet to start migrating more quickly into the inner regions of the disc.
As it migrates, it carries with it a resonant chain of six planets, all in first order resonances as shown by the inward migration of the group of planets in the bottom panel of Fig. \ref{fig:pltml_sim}.
Remaining in the outer disc, are four planets, with masses $\mpl<0.8\me$, trapped at the outer edge of a zero migration zone.

The inner convoy of six planets quickly reaches the trapped planet at the active turbulent region.
Interacting with this planet and a surrounding swarm of planetesimals after $\sim0.75$ Myr disrupts the fragile resonant chain, causing some of the planets to collide, further increasing their mass.
This is shown by the vertical dashed line labelled `B'.
Only four of the six planets survive the interactions, with the most massive planet now having a mass $\mpl\sim 1.4\me$.
The inner planet also survives and becomes trapped within the resonant chain.
Along with the more massive planet, the other planets of the chain have masses $0.4\me \le\mpl\le 0.7\me$.
After 0.9 Myr, the innermost planet is pushed beyond the active turbulent region, allowing it to migrate quickly towards the central star.
This allows the most massive planet to migrate closer to the active turbulent region, pushing another planet in past the region, before having its own migration stalled.
The two inner planets now migrate in to the inner edge of the disc, where their migration stalls after 1.2 Myr.
The other two planets in the initial chain continue to accrete neighbouring planetesimals without undergoing significant migration.
One of the planets increases its mass sufficiently, allowing its corotation torque to become stronger than its Lindblad torque, causing the planet to migrate outwards to the outer edge of a zero migration region.
As the inner chain of planets undergoes the significant dynamical interactions, the outer chain of planets slowly migrated inwards, without experiencing any significant dynamical interactions.
This is shown by the slow inwards migration of the upper group of lines in the bottom panel of Fig. \ref{fig:pltml_sim}.

\begin{figure}
\centering
\includegraphics[scale=0.4]{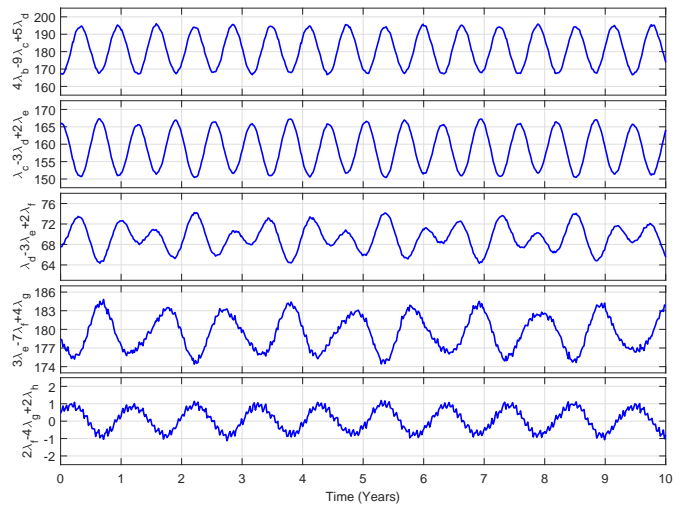}
\caption{Three body resonant angles for the planets with orbital periods < 20 d in the simulation discussed in Sect. \ref{sec:pltml_example} and shown in Fig. \ref{fig:pltml_sim}.
The timespan of 10 years is taken at the end of the simulation.}
\label{fig:pltml_3body}
\end{figure}

As the protoplanetary disc evolves, the regions where the corotation torque dominates the Lindblad torque move in closer to the central star and to lower masses.
This is due to the temperature and viscosity of the disc decreasing as the gas surface density drops, increasing the viscous and entropy time-scales of gas orbiting in the corotation region of the planet.
Since the corotation torque is at its strongest when the viscous and entropy time-scales are approximately equal to half the horseshoe libration time-scale, it is clear that as these time-scales increase, the mass of the planet's that have the strongest corotation torques decreases.
This evolution of the disc, is the reason why planets trapped in zero migration regions slowly migrate inwards as the disc evolves, rather than maintaining their positions.

After 1.8 Myr, the disc has evolved such that the most massive planet of the outer chain of planets, of mass $\mpl\sim0.8\me$, has begun to saturate its corotation torque, allowing the Lindblad torque to dominate its migration, forcing the planet to migrate inwards.
As it migrates, it carries with it the other planets of its convoy.
This can be seen by the dashed line labelled `C' and the change in slope of the upper lines in the bottom panel of Fig. \ref{fig:pltml_sim}.
The chain quickly catches up with the two planets trapped at another of the zero migration zones, pushing them past the zone, that allows them to quickly migrate in to meet the most massive planet of the system that was still trapped at the outer edge of the active turbulent region.
The outer convoy of planets then quickly migrated into the inner system, reaching the now three planet chain trapped by the active turbulent region after 2.1 Myr.
This now seven planet chain had its migration substantially slowed.
However there was enough migration to allow the chain to begin interacting with the two innermost planets of the system, forcing them to migrate in closer to the central star, resulting in the innermost planet impacting on to the star after $\sim 2.15$ Myr.

As the disc cooled below 1000 K, the active turbulent region disappeared, which allowed the chain of planets to migrate in to the inner edge of the disc, where the most massive planet became trapped, stalling the migration of those with longer periods, and forcing the innermost planet to orbit beyond the disc's inner edge, in the magnetospheric cavity.
After 2.25 Myr, the chain of planets had a significant dynamical interaction, with two of the planets colliding.
This is shown by the dashed line labelled `D' in Fig. \ref{fig:pltml_sim}, and led to the planets settling into a new resonant configuration.
The new resonant configuration also contained a second-order resonance, 5:3, between the second and third planets of the system.
All of the other planets remained in first order resonances, either 3:2 or 4:3.
Remember that we define the planets as being in resonance if their resonance angles are found to be librating.

Due to the evolution of the migration and eccentricity damping time-scales as the gas disc dissipated after 4 Myr, the resonant dynamics were slightly altered, but without creating instabilities that would change the global configuration of the system.
This is shown by the dashed line labelled `E' in Fig. \ref{fig:pltml_sim}.
For example the resonant angle between the 4th and 5th planets before the interactions was librating around $205^{\circ}$, with an erratic amplitude upto $45^{\circ}$.
After the dissipation of the disc, the amplitude of libration was reduced to just $4^{\circ}$, and was now librating around $153^{\circ}$.
This change in resonant angles occurred for most planetary pairs in the system, and for all resonant pairs, the libration amplitudes fell dramatically to small values.
The planet eccentricities also reduced in amplitude significantly as can be seen by their evolution in the middle plot of Fig. \ref{fig:pltml_sim}.
The planets remained in this configuration as the disc fully dispersed after 5.5 Myr, and also for the next 4.5 Myr, as the simulation ran to 10 Myr to allow the systems to dynamically evolve in an undamped environment.

When comparing this system to \trap, it is clear from Fig. \ref{fig:pltml_sim_mvp} that the systems look very similar in terms of planets masses and the range in periods that the planets occupy, i.e. those less than 20 days.
As discussed in the text above, all of the planets are in a resonant chain, as can be seen in Fig. \ref{fig:final_systems} where the system denoted `Planetesimal System 2' corresponds to the system described here.
In order to compare with the \trap ~system where all of the planets are shown to be in three-body resonance \citep{LugerTrappist1-h}, we grouped the planets into a chain of three-body resonances.
In Fig. \ref{fig:pltml_3body}, we show these three-body resonant angles over a 10 year period (taken at the end of the simulation), of the seven innermost planets in the simulated system described above, using standard planet nomenclature.
The precise resonant parameters are shown respectively on the $y$-axis, and it can be clearly seen that the resonant angles are librating.
The amplitudes of librations vary between the difference resonances, with some having very small librations of only a couple of degrees (bottom panel), while others have libration amplitudes of upto $25^{\circ}$ (top panel).
These libration amplitudes are consistent with those observed in the \trap ~system \citep{LugerTrappist1-h}.

When looking at the water content of the simulated planets, we find that they are extremely water rich, $\sim50\%$.
This is unsurprising, since the planets accreted the majority of their mass outside the iceline before migrating in to their final positions.
As the planetary system would evolve over Gyr time-scales, some of this water would be lost from the planets due to photoevaporation as the planets orbit extremely close to their central star \citep{Owen17,Jin18}.

As there were no late dynamical interactions in the system, i.e. after the disc had fully dispersed, the planets remained coplanar as their inclinations had been damped by the gas disc before it dispersed.
Planetary eccentricities were also small, with all of the planets having eccentricities $e_{\rm p}<0.1$.
These eccentricities could be reduced further over time through tidal damping, which would also act on the planet's semimajor axes, reducing the compactness of the system.
This could act to move the planets away from their observed first-order resonances and cause them to appear to be in higher order resonances, such as those seen in \trap ~\citep{PapaloizouTrappist}.

\section{Pebbles}
\label{sec:pebbles}
Where Sect. \ref{sec:planetesimals} formed Trappist-like systems through the accretion of planetesimals, we now consider the formation of Trappist-like systems through the accretion of pebbles on to planetary embryos.
Like the planetesimal accretion scenario above, we place numerous embryos into a protoplanetary disc at the beginning of a simulation, and then allow them to accrete pebbles from the surrounding disc.
Before presenting the results of an example simulation, we will outline the basic pebble accretion model that we have incorporated into our models.

\subsection{Accretion of Pebbles}
\label{sec:pebbles_model}
To account for the pebbles in the disc, we implement the pebble models of \citet{Lambrechts12,Lambrechts14} into our simulations.
As a protoplanetary disc evolves, a pebble production front extends outwards from the centre of the system as small pebbles and dust grains fall towards the disc midplane, gradually growing in size.
Once the pebbles that form reach a sufficient size they begin to migrate inwards through the disc due to aerodynamic drag.
The location of this pebble production front is defined as:
\begin{equation}
r_{\rm g}(t) = \left(\frac{3}{16}\right)^{1/3}(GM_*)^{1/3}(\epsilon_dZ_{0})^{2/3}t^{2/3}
\end{equation}
where $\epsilon_d = 0.05$ is a free parameter that depends on the growth efficiency of pebbles, whilst $Z_{0}$ is the solids-to-gas ratio.
Since this front moves outwards over time, this provides a constant mass flux of inwardly drifting pebbles equal to:
\begin{equation}
\label{eq:massflux}
\dot{M}_{\rm flux} = 2\pi r_{\rm g}\dfrac{dr_{\rm g}}{dt}Z_{\rm peb}(r_{\rm g})\Sigma_{\rm gas}(r_{\rm g})
\end{equation}
where $Z_{\rm peb}$ denotes the metallicity that is comprised solely of pebbles.
Combining the metallicity comprised solely of pebbles with that to which contributes to the remaining dust in the disc, gives the total metallicity of the system:
\begin{equation}
Z_{0} = Z_{\rm peb} + Z_{\rm dust}.
\end{equation}
In all of the simulations performed using pebble accretion, pebbles make up between 50 and 90 $\%$ of the total metallicity, and we assume that this ratio remains constant throughout the entire disc lifetime.
The remaining metallicity is locked up within small dust grains that contribute to the opacity of the disc when calculating its thermal structure, and again we assume this remains constant over time.
Assuming that the mass flux of pebbles originating from $r_{\rm g}$ is constant throughout the disc, we follow \citet{Lambrechts14} in defining the pebble surface density, $\Sigma_{\rm peb}$, as the following:
\begin{equation}
\Sigma_{\rm peb} = \sqrt{\dfrac{2\dot{M}_{\rm flux}\Sigma_{\rm g}}{\sqrt{3}\epsilon_p r v_K}}
\end{equation}
where $\epsilon_p$ is the coagulation efficiency between pebbles which we assume to be equal to 0.5, $r$ is the radial distance to the star, and $v_K$ is the local keplerian velocity.
As pebbles drift inwards, eventually they cross the water iceline, which we take as being where the local disc temperature is equal to 170 K.
Since pebbles are mostly comprised of ice and silicates, when they cross the iceline, the ices sublimate releasing trapped silicates, reducing the mass and size of the remaining pebbles.
To account for sublimation, we multiply the pebble surface density for radial locations interior to the iceline by a factor of 0.5 \citep{Lambrechts14}.

As the pebbles drift through the disc, they can also encounter planetary embryos and given the right conditions be accreted by the embryos.
We separate the accretion rate of pebbles into two regimes, depending on the mass of the embryo and the vertical extent of the pebbles, the pebble scale height $H_{\rm peb}$
\begin{equation}
H_{\rm peb}=H\sqrt{\frac{\alpha}{\tau_f}}
\end{equation}
where $H$ is the local disc scale height, $\alpha$ is the local viscous parameter, and $\tau_f$ is the Stokes number that depends on the physical size of the pebbles.
In the 2D regime, where the embryo's Hill radius, $r_{H} = r\sqrt[3]{m_{\rm p}/(3M_*)}$, is larger than the pebble scale height, then the accretion follows \citet{Lambrechts14}:
\begin{equation}
\label{eq:2dmdot}
\dot{M}_{\rm c,2D} = r_H v_H \Sigma_{\rm peb}
\end{equation}
where $v_H = \Omega_K\times r_H$ is the Hill speed.
In the other regime, where the embryo's Hill radius is less than the pebble scale height, then pebble accretion proceeds in the 3D mode \citep{Bitsch15}:
\begin{equation}
\label{eq:3dmdot}
\dot{M}_{\rm c,3D} = \dot{M}_{\rm c,2D}\left(\dfrac{\pi(\tau_f/0.1)^{1/3}r_H}{2\sqrt{2\pi}H_{\rm peb}}\right).
\end{equation} 
Since some of the pebbles are being accreted by the embryos, they can no longer drift further inwards.
This alters the mass flux of pebbles defined in eq. \ref{eq:massflux} to the following:
\begin{equation}
\dot{M}_{\rm flux}(r) = \dot{M}_{\rm flux} - \dot{M}_{\rm c}(r_{\rm p} \geq r)
\end{equation}
where $\dot{M}_{\rm c}(r_{\rm p} \geq r)$ sums up the mass flux of pebbles accreted by embryos exterior to an orbital radius $r$.
Embryos continue to accrete pebbles from the disc until they reach the so-called pebble isolation mass.
At this mass, an embryo begins to perturb the local disc structure, forming a significant pressure bump in the disc just exterior to its orbit, which alters the rotation velocity of the gas and halts the inward drift of pebbles, and therefore the ability for planets orbiting interior to the embryo being able to accrete pebbles.
We follow \cite{Lambrechts14} and define the pebble isolation mass as:
\begin{equation}
\label{eq:peb_iso_mass}
q_{\rm iso} = \frac{h^3}{2}
\end{equation}
where $q_{\rm iso} = m_{\rm iso}/m_*$, and $h$ is the disc aspect ratio, $H/r$.

\subsubsection{Ablation of pebbles in a planet's envelope}
\label{sec:ablation}
When pebbles plunge through a planet's envelope towards the planet's core, it is typically assumed that they do so without interacting with the envelope.
However, they do actually interact with the gaseous envelope and experience aerodynamic drag \citep{Adachi,Weidenschilling_77}.
Due to their physical size (mm--m regimes), the drag forces act to heat the pebble, causing its outer layers to vaporise and ablate \citep{Podolak88}.
Depending on the pebble properties, i.e. composition and density, and on the envelope properties, i.e. mass and density, the thermal ablation of the pebble continues until either it is fully ablated, or the remains of the pebble impacts the physical surface of the planet.

In our models we follow \citet{Alibert17}, in calculating the envelope mass required to thermally ablate pebbles that are accreted by a planet.
For solids that are the sizes of pebbles, the main ablative process that removes mass is that of ambient heating from the local gas ,i.e. the envelope of the planet.
In accounting for thermal ablation, \citet{Alibert17} provide fits to the numerical results of \cite{Mordasini06}:
\begin{equation}
\label{eq:ablation}
\dfrac{M_{\rm env}}{0.001 \me} = \left(\frac{s}{10 {\rm ~cm}}\right)^{1.17}
\end{equation}
where $s$ is the physical size of the pebble.
When taking into account that this fit was calculated for stony objects and not those abundant with water ice, we find that the envelope mass required to thermally ablate water rich objects was approximately a factor of 10 lower than that given in the equation above (priv comm with C. Mordasini).

Preliminary tests of gas accretion onto planets with core masses comparable to an Earth mass found that only meagre gaseous envelopes could be accreted throughout the lifetime of the disc.
These tests utilised the 1D self-consistent gas accretion model outlined in \cite{CPN17} and found that envelope masses never exceeded 1$\%$ of the total mass of the planet.
When the solid accretion luminosity was included in the calculations, the envelope sizes were generally much smaller than 1$\%$.
Since the envelopes are of such small mass, typically stony pebbles are still able to reach the surface of the planet when they are accreted by planets in the simulations.
For the water rich pebbles, i.e. those accreted outside the water iceline, we assume that as they undergo thermal ablation, through heating from the ambient gas, all of the water ice is lost and the remaining stony component of the pebbles continues to fall to the surface of the planet.
The water is then assumed to reside in the planet's envelope, thus changing the composition of the envelope, which can significantly alter the gas accretion rates, allowing planets to accrete gas much quicker than if the envelope was purely hydrogen and helium \citep{Venturini15}.
This change in composition should be taken into account when calculating the structure of the envelope and the corresponding gas accretion rate.
However, recent studies by \cite{Ormel15b} have shown that the envelopes of growing planets are constantly replenished by gas coming from the local protoplanetary disc, on a time-scale that depends on the properties of both the planet and the local disc.
For planets with very low mass envelopes, this results in the entirety of the envelope being constantly replenished with the local gas disc, thus restoring the composition of the envelope to being reminiscent of the protoplanetary disc, i.e. eliminating the metallicity enhancement of the envelope.
We therefore include this effect here where we assume that the water content within the planet's envelope is the same as that of the local disc material, and any enhancement due to that of ablated pebbles is simply recycled back into the local gas disc.

However, recent studies by \cite{Ormel15b} have shown that the envelopes of growing planets are constantly replenished by gas coming from the local protoplanetary disc, on a time-scale that depends on the properties of both the planet and the local disc.
One caveat with this work however, is that their simulations were only isothermal and may not have captured all of the physics in the envelope.
More recently \citet{Lambrechts17}, using a more realistic energy transfer model for the envelope, found results that are consistent with \citet{Ormel15b}, in that there is significant recycling within a growing planet's envelope.
They also found that the envelope was generally undergoing recycling with the local gas disc, down to the radiative-convective boundary, also finding that the water iceline was situated within the convective region of the planet if the planet was orbiting exterior to the iceline of the disc.
This would indicate that the ablated water ice would become trapped in the convective layers of the envelope and would undergo minimal recycling with the local gas disc.
In other work by \cite{Kurokawa18}, who use beta cooling models, and neglect any luminosity generated through the accretion of solids, find that there is a buoyancy barrier that acts against the recycling of the envelope.
However, the extent to which recycling is reduced is found to be dependent on the strength of the beta cooling used.
They also find that in adiabatic cases, the envelope undergoes significant recycling.
More recent work by \cite{Bethune19b}, again using isothermal 3D simulations found that recycling of the envelope can penetrate all the way to the planet's core, except for larger mass cores where shocks act to prevent recycling.

Whilst there are many works with different results when examining the recycling of a planet's envelope, they all indicate that recycling is more efficient for lower-mass cores.
Given that the typical planet masses used in the works described above are of super-Earth mass, i.e. $\mpl>3\me$, it is reasonable to assume that the recycling of meagre envelopes around sub-terrestrial mass planets is extremely efficient and may reach down to the planet's surface \citep{Bethune19b}.
In addition, when calculating 1D envelope structure models of these small envelopes \citep{CPN17}, the radiative-convective boundary is also very close to the core of the planet, indicating that recycling will go deep into the planets' envelope \citep{Lambrechts17}.
Therefore it is reasonable to assume, given that the core and envelope masses in our simulations are small, especially those of the envelope, that the recycling is efficient and any envelope attained in our simulations is fully recycled by the protoplanetary disc.
This would also allow for the water lost to ablation as it crossed the water iceline in the convective layer of the planet's envelope \citep{Lambrechts17}, to also be recycled back into the protoplanetary disc,without the need to account for its effects on the composition of the planet's envelope \citep{Venturini15}.

For our models we assume that the envelope mass of a planet is equal to $0.1\%$ of the planet's total mass.
Preliminary tests found that planets could retain envelopes of this size when accreting pebbles at appropriate rates.
If this mass is greater than the mass required for ablation in eq. \ref{eq:ablation}, then we assume the pebble has ablated and the water content is lost to the disc, and the solid component is accreted by the planet.

\begin{figure*}
\centering
\includegraphics[scale=0.8]{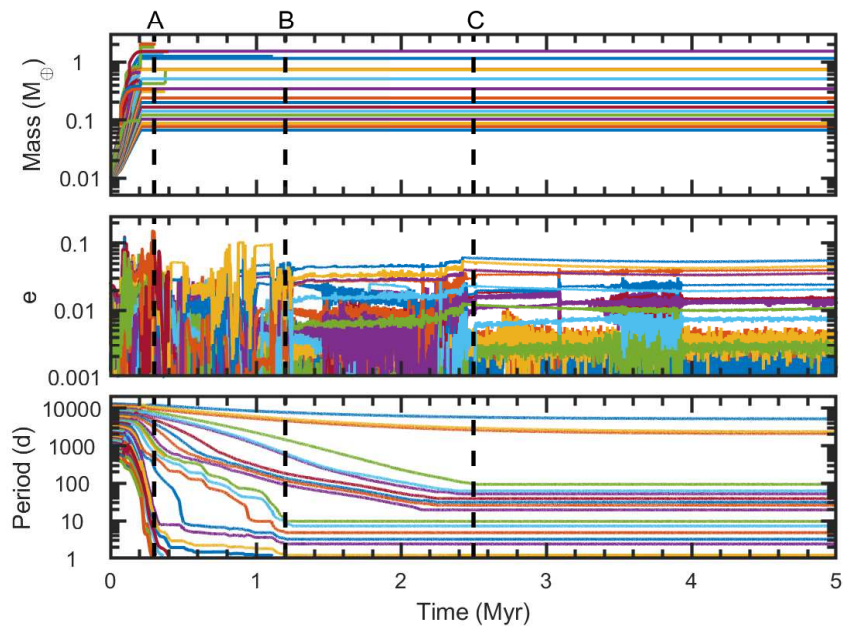}
\caption{Temporal evolution of planet masses (top), eccentricities (middle) and periods (bottom).
The noted letters indicate times of interesting events that are described in the text.}
\label{fig:peb_sim}
\end{figure*}

\subsection{Initial Setup}
For these simulations we place 40 planetary embryos, of mass 0.01 $\me$ and semimajor axes between 0.5 and 5 $\au$, into a protoplanetary disc such as that described in Sect. \ref{sec:models}.
The mass chosen for the start of the pebble accretion simulations is comparable to the transition mass, where the pebble accretion rate transitions from the Bondi regime to the Hill regime \citep{Bitsch15}.
We note that the initial embryo masses are different to those for the planetesimal accretion scenario above, however the time taken for embryos to increase their mass in the pebble accretion scenario from 0.01 $\me$ to 0.1 $\me$ is short compared to the lifetime of the disc, and also dynamical and migration time-scales.
Conversely, for the planetesimal accretion scenario, we find that the time to reach reach an initial embryo mass of $0.1\me$ from $0.01\me$ would take at most 0.5 Myr for the discs in our simulations, depending on the location in the disc and the density of planetesimals.
The main reasons for the differences in initial embryo masses, is computational constraints on the side of planetesimal accretion (starting at such a low embryo mass would require the inclusion of the $>10,000$ planetesimals that would result in extremely long computational run times), and for the initial embryo mass in the pebble accretion scenario being similar to the transition mass so as not to shock the system and have the embryos accreting too quickly, and reaching the pebble isolation mass almost instantaneously.
Ideally we would start both sets of simulations at even smaller embryo masses, i.e. similar to the most massive planetesimal that may arise from the streaming instability \citep{Abod19}, but this again would increase the computational run time in both scenarios, and would add further parameters to the models, as both scenarios would then include a planetesimal formation mechanism.

Initial eccentricities and inclinations are randomised according to a Rayleigh distribution, with scale parameters $e_0=0.01$ and $i_0=0.25^{\circ}$, respectively.
Collisions between planets are assumed to result in perfect mergers.
We use initial disc masses of between 2.7 and 8 $\%$ of the mass of \trap.
For the conversion efficiency of dust to pebbles, we take efficiencies of 50:10:90 $\%$.
To account for different disc lifetimes, we modify the photoevaporation rate such that the $f_{41}$ factor, representing the number of ionising photons emitted by the central star per second, takes values of either: 0.0001, 0.001, 0.01 or 0.1.
This mainly affects the evolution of the disc at the the end of its lifetime, as it determines when the disc begins to quickly disperse from the inside-out.
We also account for the ablation of pebbles in a planet's primordial envelope in one set of simulations and ignore it in another.
Finally, we run two versions of each set of parameters, where the planet positions and velocities are initialised using a different random number seed.
We ran all simulations for 10 Myr which accounts for the whole disc lifetime, and also allows some time for the systems of formed planets to continue to evolve through scattering and collisions in an undamped environment after the gas disc has fully dispersed.

\subsection{Example Simulation}
\label{sec:peb_example}
\begin{figure}
\centering
\includegraphics[scale=0.6]{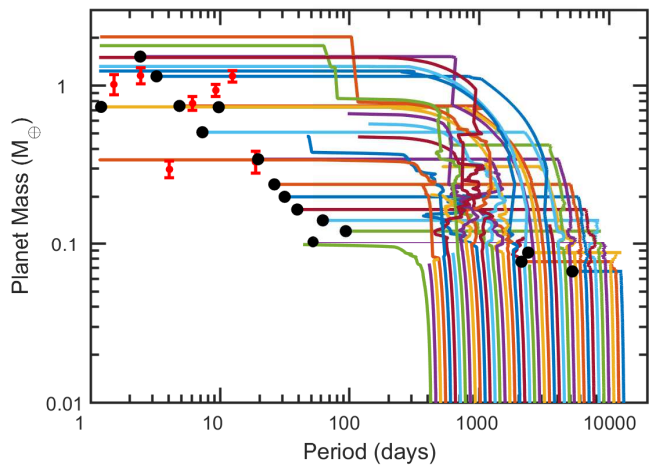}
\caption{Evolution of planet mass versus period for an example simulation.
Filled black circles represent final masses and periods for surviving planets.
The red dots indicates the periods and masses of the \trap ~planets with their appropriate error bars \citep{Grimm18}.}
\label{fig:peb_sim_mvp}
\end{figure}

We now show the results of a single simulation, representative of the evolution of a Trappist-like system.
The evolution of planet masses, eccentricities and periods are shown in Fig. \ref{fig:peb_sim}.
Figure \ref{fig:peb_sim_mvp} shows a mass versus period representation of the system, where black dots represent the simulated planets final masses and periods, and red points represent the \trap ~planets with appropriate errorbars.

Looking at the temporal evolution of the system in Fig. \ref{fig:peb_sim}, it is clear to see that most of the accretion occurs at the start of the simulation, within the first 0.2 Myr.
From the beginning of the simulation, the pebble growth front quickly passes the small planetary embryos, allowing them to begin accreting pebbles.
As the pebble front advances outwards, a number of planetary embryos reach masses where they begin to significantly migrate within the disc.
This can be seen in the bottom panel of Fig. \ref{fig:peb_sim} where the planets begin to migrate towards the central star.
The arcing of the tracks when the planets reach $\mpl \sim0.5\me$ in Fig. \ref{fig:peb_sim_mvp} also shows where migration is beginning to take hold.
Accretion of pebbles stops after $\sim$ 0.2 Myr, as shown by the planet tracks going horizontal in the top panel of Fig. \ref{fig:peb_sim}.
The stopping of accretion is due to the pebble growth front reaching the outer edge of the disc, reducing the mass flux of pebbles to zero.
At the end of accretion phase, mutual collisions between planets orbiting between $0.1$--$0.5 \au$, results in a number of $\sim$ Earth-mass planets forming.
These planets then quickly migrate to the inner edge of the disc, along with a couple of less massive planets that are forced to migrate as part of two resonant chains.
This can be seen in Fig. \ref{fig:peb_sim} at the vertical dashed line labelled `A'.
As the first chain of planets reaches the inner edge of the disc, the planets are held up by the active turbulent region at $P\sim 10$ d, where the increase in turbulence enhances their corotation torques, allowing them to balance the Lindblad torques, creating a planet trap.
This trap only acts as a brief respite for the inner chain of planets though, as the outer chain of planets migrates in closer to the inner chain, then pushes the inner chain of planets past the active turbulent region.
The inner chain then gets trapped at the inner edge of the disc, until  a number of the outer chain of planets pushes them past the inner edge, forcing them to impact on to the central star.
The outermost planet of the outer chain of planets becomes trapped at the outer edge of the active turbulent region, with the rest of the planets in the outer chain being pushed past this trap,
After 0.4 Myr, only three planets of the two chains of planets remains (12 originally), the other planets lost either to collisions with more massive planets or to the central star.

After a further 0.1 Myr, a fourth planet with mass $\mpl\sim 1\me$ has migrated to the inner regions of the disc, entering into resonance with the chain of remaining planets mentioned above.
These planets then migrate inwards slowly as the disc cools and the active turbulent region moves closer to the central star.
After 1 Myr, another planet migrated inwards and joined the inner grouping of planets.
Shortly after another two planets of masses $0.5 \me$ and $0.7 \me$, migrate in towards the star and join the outer edge of the group of planets.
These planets then push the chain of planets in further, causing the innermost planet to impact on to the central star.
As of 1.3 Myr, 6 planets now orbit in a resonant chain, with periods out to 10 d, and are shown as the inner 6 planets on the left panel of Fig. \ref{fig:peb_sim}.
This can be seen by the vertical dashed line labelled `B' in Fig. \ref{fig:peb_sim}.
After 2.5 Myr, another 7 low mass planets had migrated in to join the chain of planets with the innermost of these 7 planets entering into a 2:1 resonance with the outermost planet of the inner chain.
This can be seen at the vertical dashed line labelled `C' in Fig. \ref{fig:peb_sim}.
The slowness of their migration was due to their low mass, with many being around a Mars mass ($0.1 \me$).

The three remaining sub-Mars mass planets in the outer regions of the disc continue to migrate in slowly as the disc evolves.
After 5 Myr, the disc had undergone full disc dispersal, leaving the planets to evolve in an undamped environment for the remaining 5 Myr of the simulation.
However, due to all of the planets being in stable resonant chains, there was no further evolution of the planets dynamically, and the final state of the simulation is shown by the black dots in Fig. \ref{fig:peb_sim_mvp}.

\begin{figure}
\centering
\includegraphics[scale=0.4]{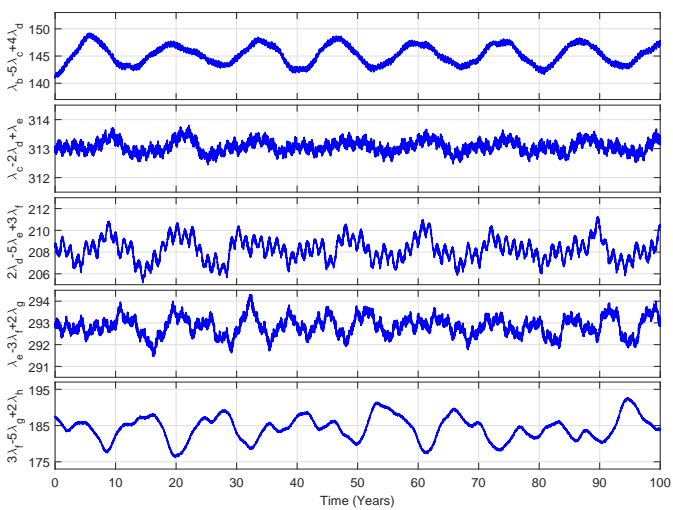}
\caption{Three body resonant angles for the planets with orbital periods < 20 d in the simulation discussed in Sect. \ref{sec:peb_example} and whose evolution is shown in Fig. \ref{fig:peb_sim}.
The timespan of 10 years is taken at the end of the simulation.}
\label{fig:peb_3body}
\end{figure}

\begin{figure*}
\centering
\vspace{1cm}
\includegraphics[scale=0.7]{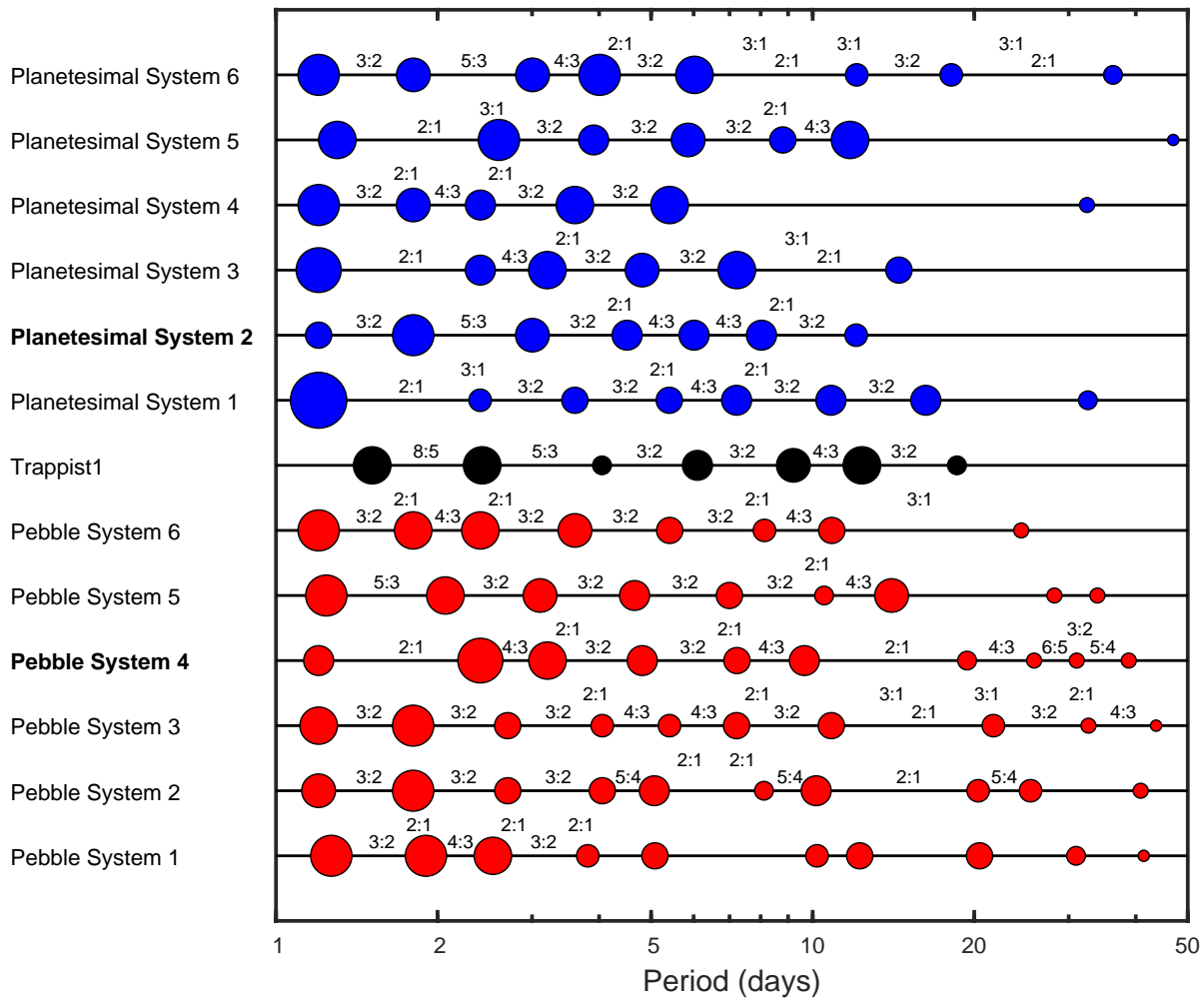}
\caption{Simulated Trappist-like systems from the planetesimal accretion scenario (top 6 rows) and the pebble accretion scenario (bottom 6 rows).
Periods are shown on the $x$-axis and planet masses are indicated by the symbol size.
Resonances are also shown between neighbouring planets and when relevant, between non-neighbouring planets.
The systems highlighted in bold represent those described in Sects. \ref{sec:pltml_example} and \ref{sec:peb_example} for the planetesimal and pebble accretion scenarios respectfully.
The \trap ~system is shown in between the two scenarios for comparison.}
\label{fig:final_systems}
\end{figure*}

When comparing this system of planets to \trap , it is clear from Fig. \ref{fig:peb_sim_mvp} that the system looks similar in terms of the planet masses and the period range they occupy, i.e. upto 20 days.
In the simulated system there are also more planets orbiting both with slightly longer periods (upto 100 days), and also some orbiting far from the central star.
It is seen in this simulation that all of the inner planets are in two-body resonance, and consequently in three-body resonance, again similar to the \trap ~system \citep{LugerTrappist1-h}.
The two-body resonances for the simulated system can be seen in Fig. \ref{fig:final_systems} where the system discussed here corresponds to the system `Pebble System 4' there.
As all of the planets are in a resonant chain, we determine the three-body resonant angles, where the inner 7 planets are shown in Fig. \ref{fig:peb_3body}, with the orbital data being taken over a 100 year period at the end of the simulation.
The precise resonant parameters are shown on the $y$-axis labels of each respective panel, where we again use standard nomenclature.
They are all clearly seen to be in three-body resonance, with librations of upto 10 degrees, comparable to the \trap ~planets.

Since there were no late dynamical interactions in the system, these planets remained coplanar as their inclinations had been damped during the disc lifetime.
The inner planets also retained very low eccentricities, $e_{\rm p}<0.05$.
Tidal damping could act to reduce these eccentricities, further stabilising the system.
The effects of tides could also act to force the inner planets to migrate closer to central star, breaking the appearance of first-order two-body resonances.
Since the planets are in three-body resonance, this could possibly be retained as the planets migrate away from the two-body resonance equilibrium \citep{PapaloizouTrappist}.

\section{Comparisons}
\label{sec:comparisions}

Sections \ref{sec:pltml_example} and \ref{sec:peb_example} detailed example simulations of the planetesimal and pebble accretion scenarios that produced a system that looked similar to \trap.
We now look at all of the simulations from both scenarios as a whole, initially comparing them to how well they form systems similar to \trap, and then we examine how similar the systems from each scenario are to each other.

\subsection{Similarity Criterion}
\label{sec:comp_similarity}
In order to compare the simulated systems to that of \trap ~we create a similarity criterion that compares the outcomes of models against their observed values.
This comparison can be performed `by eye' (e.g. looking at how similar a range of outcomes predicted by a model are similar to the observed values), but this can be very subjective.
Note that this comparison does not confirm the validity of the models, since they can be based on wrong physical assumptions, it merely allows for the results to be compared with observations.
The use of a similarity criterion is therefore no solution to this problem of validity, but is rather a way to make such a comparison objective.
Indeed, when comparing for example the mass distributions predicted by two models with that of observed planets, it is common to draw histograms of the three quantities.
One first approach is then to just `look' at the histograms and decide whether or not they `look the same' or not.
In our example with two models, one can then decide that the histogram from one model `looks like' the observed one more than that from the other model.
Such a comparison is in general rather subjective, in particular when both models produce not totally different outcomes.
A second, more objective, approach is to make a statistical test (e.g. KS test for cumulative histograms) which quantifies how similar the histograms are.
Then one can quantify to which extent the histograms from either model `look more like' that from observations.
Again, this does not prove that one model is better or more accurate than the other, since, as we said above, one or both models could be intrinsically wrong (e.g. they could be based on equations that do not conserve mass).
Therefore the similarity criterion we propose should be seen as the counterpart of a KS test in the case of histograms.
In our case, comparing the outcome of one model with \trap ~and the outcome of another model with \trap ~means comparing two sets of points in a 2D space.
Such a comparison is in general quite subjective and our similarity criterion is a way to quantify this difference.

By taking into account the overall system quantities from the \trap~ planetary system, e.g. total planet mass, we can then use the basis for a similarity criterion described above to compare those values with the synthetic planetary systems.
In theory this similarity criterion should then be able to determine the most Trappist-like synthetic planetary systems, which could inform on the formation of \trap.
As models develop, such a similarity criterion can be added to, further improving its accuracy.
Recently a similar approach of comparing planetary systems was presented in \citet{Alibert19}, of which there is a strong correlation between their results, and the results of the similarity criterion presented here, when comparing synthetic planetary systems to \trap.

The parameters that we use to calculate the similarity criterion between the synthetic planetary systems and the \trap~ planetary system are shown in Table \ref{tab:trappistcomparison}.
These parameters give a good measure of the dynamics in the system, through the total mass of the planets, their mean period ratios, and their eccentricities.
The use of the mean period ratios and their respective variance, also gives a measure of the compactness of the systems, and the presence of resonances, but does not explicitly include the resonant features of the system.
Further improvements to the model presented in \citet{Alibert19}, including the presence of resonances between neighbouring planets, will be used to increase its effectiveness in comparing synthetic to observed planetary systems in future work (Alibert et al in prep).

Though the radius of each planet is by far the most accurate and reliable observation of the planets orbiting \trap, converting from mass to radius for the simulated systems adds a number of extra model parameters.
Mass-radius relations generally depend on a number of factors, including the mass and water content of the planet, the location of the core-mantle boundary, the fraction of heavy elements, and the compositional make-up of the heavy element fraction, and also the choice of equation of state to use for each element \citep{Wagner11,Dorn15,Hakim18}.
The choice of values for the model parameters for the planets in our simulations can give uncertainties of upto 50$\%$, making the calculated planetary radii unsuitable for comparison.
Therefore given the uncertainty in the radii of the simulated planets, and given that the mass uncertainty of the \trap ~planets from TTVs are only $\sim 10 \%$, we choose to compare the masses of the planets and not the radii within the similarity criterion.

To calculate the similarity criterion for each $i^{\rm th}$ system to \trap, we use the following:
\begin{equation}
\label{eq:distance}
d_{i}^2 = \sum^N_{j=1} (x_{j,i}-t_{j})^2
\end{equation}
where $d$ is the similarity criterion, $j$ is one of the variables, $N$ is the total number of variables and
\begin{equation}
\label{eq:distance2}
x_{j,i} = \dfrac{y_{j,i}-{\rm min}(y_{j,1:n})}{{\rm max}(y_{j,1:n}-{\rm min}(y_{j,1:n}))}
\end{equation}
and
\begin{equation}
\label{eq:distance3}
t_{j} = \dfrac{{\rm trap}(j)-{\rm min}(y_{j,1:n})}{{\rm max}(y_{j,1:n}-{\rm min}(y_{j,1:n}))}.
\end{equation}
Here the array, $y$, contains the values for the variables shown in Table \ref{tab:trappistcomparison}, whilst the `trap' array contains those values for the \trap ~planetary system, shown in the second column of Table \ref{tab:trappistcomparison}, and $n$ is the total number of systems

\begin{table}
\centering
\caption{The parameters used to deter.mine the similarity criterion given by eqs. \ref{eq:distance}--\ref{eq:distance3} that shows how similar a synthetic system is to the observed \trap ~planetary system.
The values for \trap ~are taken from \citet{Grimm18}.}
\begin{tabular}{lc}
\hline
Parameter & \trap ~\\
\hline
Mean mass & 0.8079 $\me$\\
Variance mass & 0.362 $\me$\\
Total mass & 5.655 $\me$\\
Mean period ratio & 1.5226\\
Variance period ratio & 0.1117\\
Mean eccentricity & 0.0063\\
Variance eccentricity & 0.0025\\
$\rm dM_p/dP$ & -0.0206 $\rm \me/d$\\
$\sigma(\rm M_p-M_{cp})$ & 0.3393 $\me$\\
\hline
\end{tabular}
\label{tab:trappistcomparison}
\end{table}

\subsubsection{Similar Systems to \trap}

Figure \ref{fig:final_systems} shows the simulated systems that are most similar to \trap, i.e. those with the smallest similarity criteria given by eq. \ref{eq:distance}.
The size of the markers in Fig. \ref{fig:final_systems} are proportional to the mass of the planet.
First and second order mean motion resonances are shown between neighbouring planets, and also in some instances, important first order mean motion resonances between non-neighbouring planets are also shown.
The top six systems show the most similar systems to \trap ~that formed in the planetesimal accretion scenario, whilst the bottom six systems show the most similar systems from the pebble accretion scenario.
The \trap ~planets are shown in between the two scenarios.

It is clear from Fig. \ref{fig:final_systems} that there is good similarity between the simulated systems and that of \trap.
In both scenarios, the periods that the simulated planets possess, and also the masses of the planets, are very similar to those seen in \trap.
The abundance of resonances, particularly first-order resonances, are also clear to see in the simulated systems from both scenarios, with most planets in resonance with their neighbours.
Not only are first-order resonances observed between planets, but also the second-order 5:3 resonance.
We find in the simulations, that these second-order resonances form when planets in first-order resonance embedded in the disc interact and collide, allowing the surviving planets to then migrate slowly into the 5:3 resonance.
Since they retain some eccentricity from the interactions and collisions, they are able to become trapped in the 5:3 resonance instead of migrating through the resonance, and becoming trapped in the first-order 3:2 resonance.
This is similar to what occurred in the example simulation described in Sect. \ref{sec:pltml_example}, where that final system is shown as `Planetesimal System 2' in Fig. \ref{fig:final_systems}.

It is also interesting to note that some of the planets may appear to be in higher order resonances, e.g. 8:5.
An example of this is the 5th and 6th planets of the system denoted `Pebble System 2' (second from the bottom in Fig. \ref{fig:final_systems}).
Typically higher-order resonances are extremely weak and difficult to attain in simulations involving migration, since planets normally migrate through the resonance.
However the two planets in the simulation mentioned have a period ratio close to 8:5 and given the abundance of resonances in the system, it would be expected that they would also be in resonance.
But, on closer inspection of this planet pair, we find that instead of being in the 8:5 resonance, both planets are actually in first-order resonance with a third planet, closer in towards the star.
These resonances being 5:4 and 2:1 respectively, with the non-neighbour 2:1 resonance being shown slightly above the neighbouring 5:4 resonances in Fig. \ref{fig:final_systems}.
Instead of the outermost planet migrating past the 8:5 resonance with the middle planet, it becomes trapped in the first order 2:1 resonance with the innermost planet, and now migrates as part of a chain.
It is only that the middle planet is in a 5:4 resonance with the inner planet, that the 8:5 period ratio between the middle  and outermost planet appears, and the planets are only weakly in resonance, if at all.
This explanation is consistent for the other planets seen in the 8:5 resonance in the simulations. 

\subsubsection{Comparing all systems to \trap}

\begin{figure*}
\vspace{0.5cm}
\centering
\includegraphics[scale=0.5]{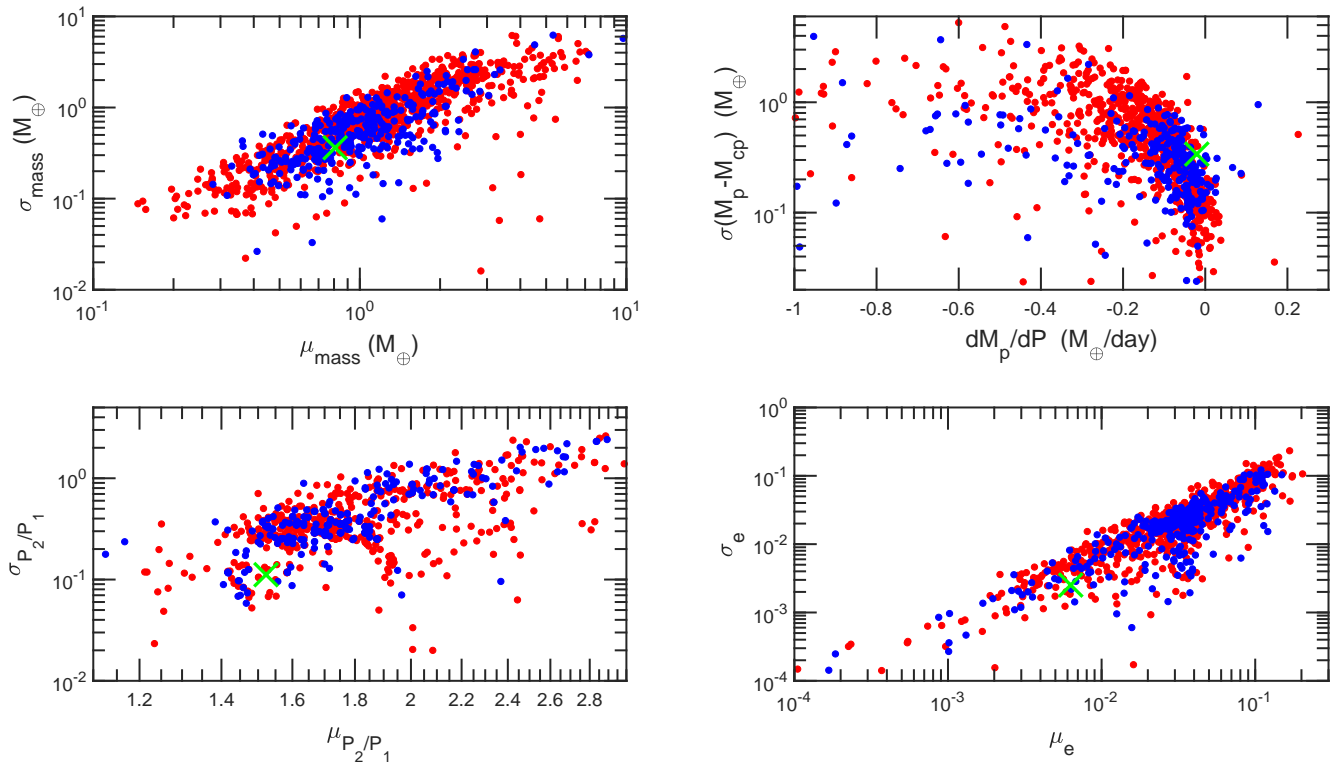}
\caption{Different components contributing to the similarity criterion (eq. \ref{eq:distance}) for systems formed by pebble accretion (red points) and planetesimal accretion (blue points).
The green crosses denote the respective values for the \trap ~system.
{\it Top left panel}: Mean planet mass versus dispersion of the planet mass.
{\it Bottom left panel}: Mean period ratios between neighbouring planets versus their dispersion.
{\it Top right panel}: Assuming that the gradient of mass as a function of period follows $y=mx+c$, the panel shows the average gradient as a component ($d\mpl / dP$), along with the dispersion of masses compared to their expected masses using the calculated straight line equation.
{\it Bottom right panel}: Mean planet mass versus dispersion of the planet mass.
All components only take into account planets with periods less than 20 days.}
\label{fig:comp_diffsims}
\end{figure*}

Figure \ref{fig:final_systems} showed the most \trap -like systems from both the pebble and planetesimal accretion scenarios, but it is interesting to see how all of the systems fare in this comparison.
The similarity criterion (eq. \ref{eq:distance}) contains numerous components to calculate the likeness of the simulated systems to \trap, see Table \ref{tab:trappistcomparison}.
These components include the average mass and eccentricities of planets with periods less than 20 days, along with their dispersions.
The average period ratios are also used, along with their dispersion values.
The other component of note, is assuming that the gradient of mass as a function of period follows $y=mx+c$, the similarity criterion including the average gradient as a component ($d\mpl / dP$), along with the dispersion of masses compared to their expected masses using the calculated straight line equation.
Figure \ref{fig:comp_diffsims} shows these components in four panels.
The top left panel shows the average mass in each system versus its dispersion.
The bottom left shows the average period ratio and its dispersion values, whilst the top right shows the mass gradient as a function of period along with its calculated dispersion values discussed above.
Finally the bottom right shows the average eccentricities along with their dispersions.
The red points show systems from the pebble accretion scenario, whilst the blue points show systems from the planetesimal accretion scenario.
The green cross shows the respective values for the \trap ~system.

When comparing the average planet mass and the dispersion in planet masses in each system upto 20 days (top left panel), we see that the pebble and planetesimal accretion scenarios overlap very well.
The majority of systems contain planets with average masses between 0.5--2 $\me$, with dispersions of between 0.1--1 $\me$.
This shows that even though the accretion of solids occurs on different time-scales, the average amount of solids accreted by planets is relatively similar, whilst also leading to similar dispersions in planet mass.
The average mass and dispersion of the \trap ~planets also compares nicely with the two populations, embedded amongst the blue and red points.

The bottom left panel of Fig. \ref{fig:comp_diffsims} shows the average period ratios between neighbouring planets, along with their respective dispersions.
Whilst there appears to be little difference between the pebble and planetesimal accretion scenarios, the red and blue points overlap quite well, there is significant variation in the observed mean period ratios.
This variation can be seen with most of the points ranging between 1.4 (between the 3:2 and 4:3 mean motion resonances) and 2.2 (exterior to the 2:1 resonance), showing that there is a lot of diversity in the compactness of the planetary systems.
Unsurprisingly, the more compact systems, i.e. those with small mean period ratios, contain a higher number of planets than the less compact systems.
The \trap ~system is relatively compact, with an average period ratio of 1.52 between its seven planets.
It also has a small dispersion, showing that most of the planets have period ratios close to the mean.
Interestingly this system does sit amongst a number of systems formed through planetesimal or pebble accretion, where the number of planets in these systems tends to be large, i.e. 5 or more, similar to \trap.

The top right panel shows the mass gradients for all planets within 20 days of each system, i.e. how the mass of the planets in the system changes as a function of period.
For example, if a system had a very massive planet close to the central star, and ever less massive planets as the period increased, then the gradient would be negative.
Typically for resonant chains where the migration is dominated by the more massive planets, the gradient would be closer to zero, if not positive, since the mass would be increasing as a function of period.
In fact the predicted mass gradient in the \trap ~system is only -0.02, according to the values of \citet{Grimm18}.
When comparing the systems formed through pebble accretion to those formed through planetesimal accretion, we find that they have similar gradients in mass as a function of period.
Generally the systems have slightly negative gradients, with slightly more massive planets close to the central star, and less massive planets further out.
Whilst the two populations have considerable overlap, it is interesting to note that the systems formed through pebble accretion seem to be more varied in the range of mass gradients observed in the systems.
The pebble systems are also more likely to have a wider range of masses as shown by their increased dispersions when comparing to the systems formed through planetesimal accretion.
It is also interesting to see that the calculated values for the \trap ~system overlay the two populations, again making it difficult to ascertain a preferred formation route for the \trap ~planets.

In terms of the eccentricities and inclinations of the planets in systems formed through pebble or planetesimal accretion, they are rather similar as is shown in the bottom right panel of Fig. \ref{fig:comp_diffsims} for the eccentricity case (inclinations follow a similar distribution).
This is unsurprising that there is little difference between the scenarios, since the dynamics between planets and the damping of eccentricities and inclinations from the disc, are independent of the accretion mechanism when the final masses of the planets and their distributions are similar as is shown in the top left panel of Fig. \ref{fig:comp_diffsims}.
Interestingly, the calculated eccentricities of the \trap ~planets \citep{Grimm18}, are located near to some systems formed through pebble or planetesimal accretion.
This agreement however, is at lower eccentricities than the bulk of systems for eccentricities.
Since the planets of \trap ~have had Gyr to evolve, it could be expected that some of their eccentricities and inclinations would have been damped through tidal interactions with \trap, something which has not been taken into account with the simulated systems here.
It would be prudent to assume then, that if the tidal interactions on Gyr time-scales was taken into account for the simulated systems, then the average eccentricities of the planets would decrease, bringing the bulk of planetary systems formed into greater comparison with \trap.

In general when comparing the systems formed through pebble accretion to those from planetesimal accretion, Fig. \ref{fig:comp_diffsims} shows there is very little that differentiates them.
For most parameters, the two populations seem indistinguishable.
The only parameter where there appears to be a difference is the gradient of planet mass in a system as a function of orbital period.
Systems formed through both pebble or planetesimal accretion tend to have planet mass decreasing as a function of period, but for the systems formed by pebble accretion, this gradient is typically stronger, i.e. planet mass drops off quicker as the period increases.
It is also interesting to note that for all of the parameters, the respective values for the \trap ~system overlaps both systems formed through pebble accretion, and through planetesimal accretion.
This apparent agreement between the \trap ~system and the accretion scenarios makes it difficult to determine the preferred formation mechanism of the \trap ~system.
In the next section, we will compare the two scenarios to each other, attempting to find observable differences that could help determine which scenario is favourable for the formation of planetary systems around low-mass stars.

\begin{figure*}
\centering
\includegraphics[scale=0.55]{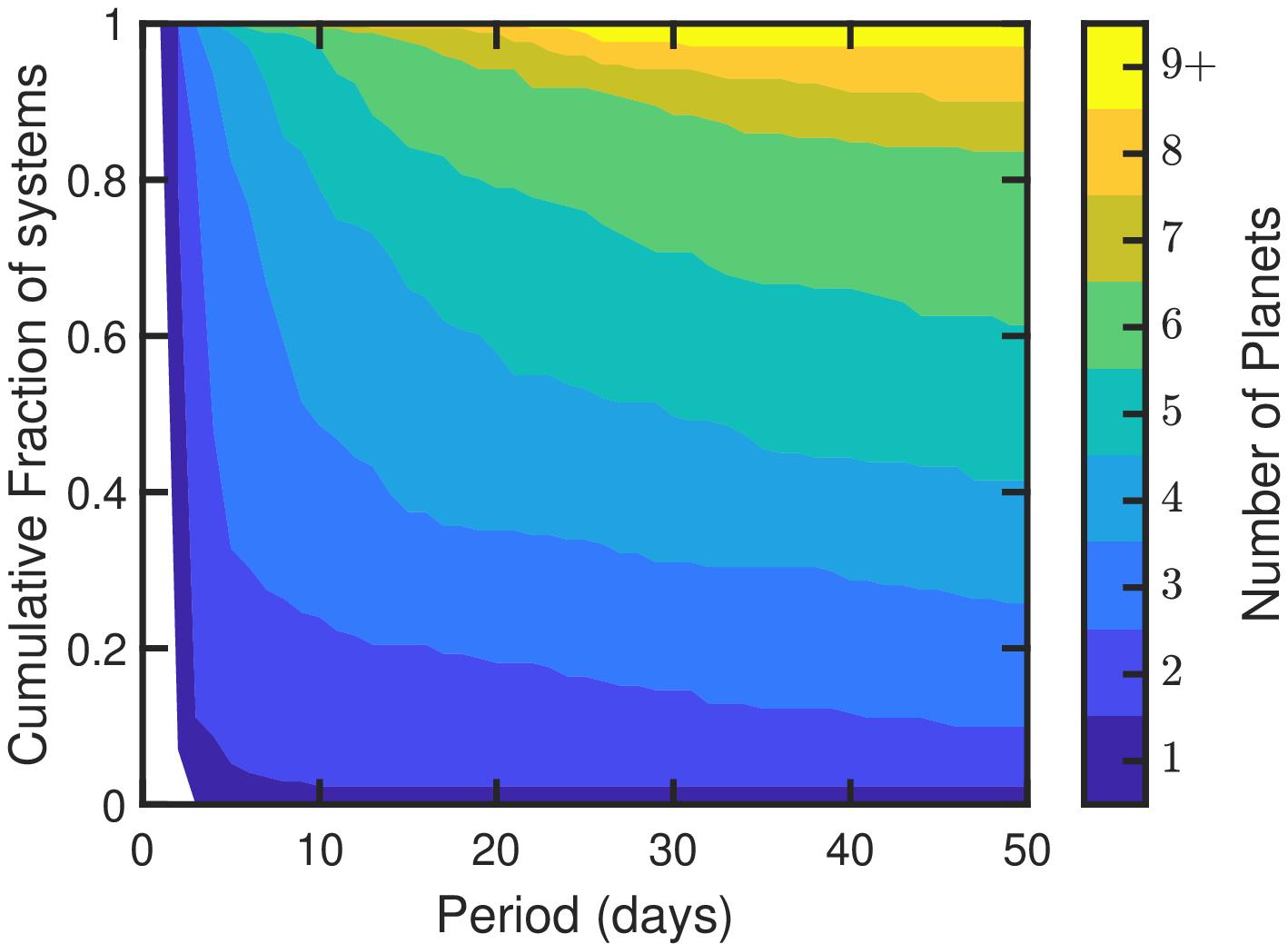}
\hspace{1cm}
\includegraphics[scale=0.55]{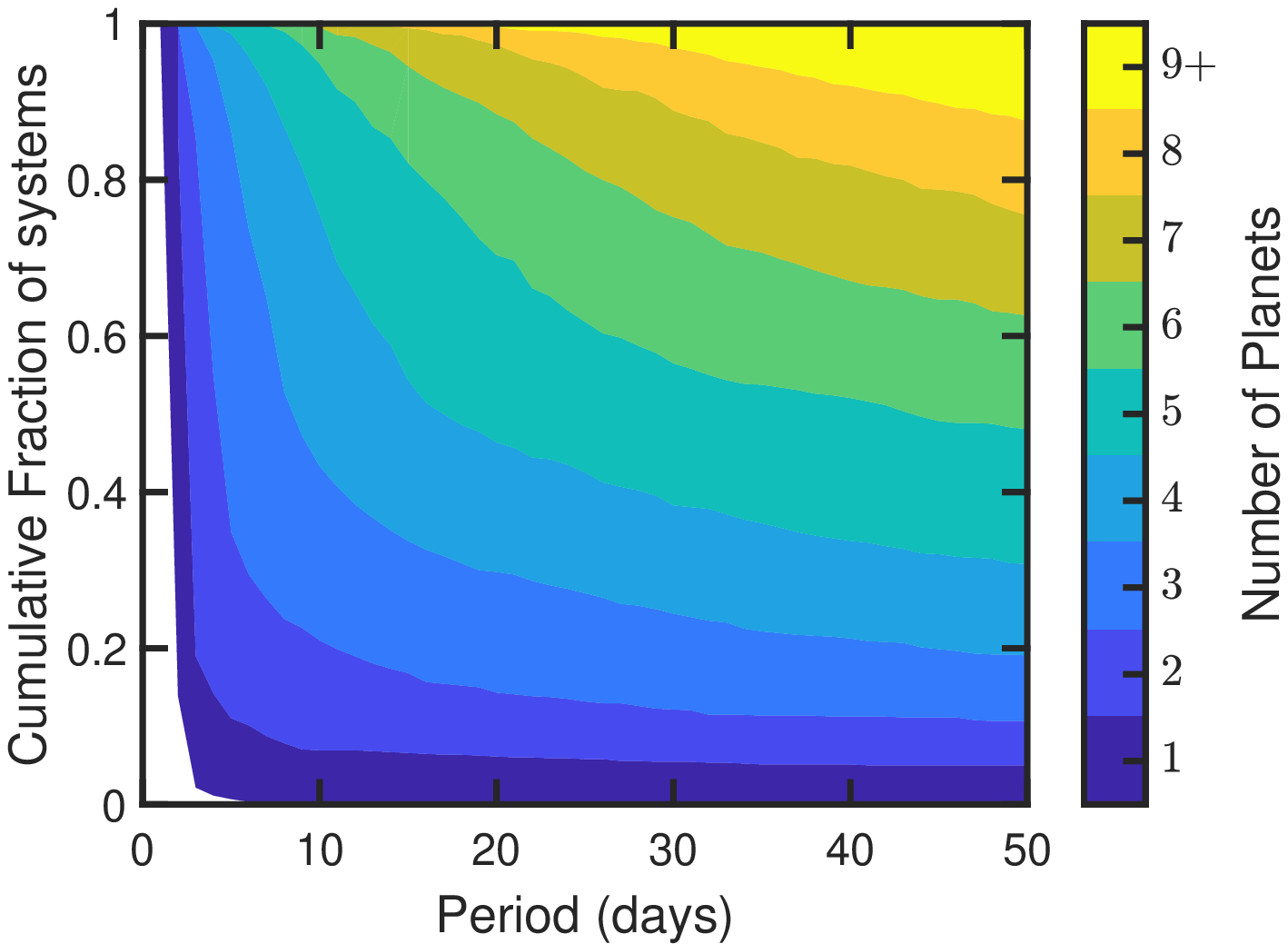}
\caption{The fraction of systems containing $n$ planets within a certain orbital period for the planetesimal accretion scenario ({\it left panel}) and for the pebble accretion scenario ({\it right panel}).
The number of planets is shown by the colour and the minimum planet mass counted was 0.1$\me$.}
\label{fig:system_fraction}
\end{figure*}

\subsection{Comparison Between the Overall Outcomes of the Pebble and Planetesimal Accretion Scenarios}

When comparing the most Trappist-like systems from each scenario to each other, it is clear to see that both scenarios easily form systems that are similar to \trap, and therefore equally similar to each other.
This similarity holds in terms of the masses, eccentricities, and inclinations of the planets, their orbital periods, the abundance of first-order mean motion resonances, the appearance of second-order mean motion resonances, and the existence of three body resonances within the systems.
In comparing to \trap, it is difficult to determine which scenario is favourable for its' formation.
Therefore we will now compare the results of the two scenarios to each other, to determine if there are any differences that might be observable.

\subsubsection{General Structure of Simulated Systems and their Observability}

As we look at all of the systems formed instead of just Trappist-like systems, it is interesting to determine what kind of systems we should expect to observe.
In determining this, Fig. \ref{fig:system_fraction} shows the cumulative fraction of systems that contain $N$ planets within a specific period, e.g. how many systems contain 7 planets, with all planets having periods less than 20 days (\trap).
The left panel of Fig. \ref{fig:system_fraction} shows the cumulative fraction for the planetesimal accretion scenario, and the right panel shows the cumulative fraction for the pebble accretion scenario.

In terms of the number of planets in a system as a function of orbital period, systems formed through pebble accretion look similar to those formed through planetesimal accretion.
For example, when looking at systems formed through pebble accretion that contain at least 5 planets, we can see that 24$\%$ of systems contain 5 or more planets within a period of 10 days, compared to 20$\%$ for those systems formed through planetesimal accretion.
Dynamically, these planets are all in stable orbits, typically in mean motion resonance with each other, see for example the systems shown in Fig. \ref{fig:final_systems}.
In fact, the first 5 planets of the \trap ~system all have periods less than $\sim9$ days.
This percentage rises to 54$\%$ for the pebble accretion systems with 5 or more planets within 20 days, and 69$\%$ within 50 days, comparable to 40$\%$ within 20 days, 54$\%$ and within 50 days for the planetesimal systems.
As we increase the number of planets in a system, these percentages obviously decrease.
So if we look at 7 planet systems, i.e. similar to \trap, we find that 12$\%$ of systems formed through pebble accretion contain 7 or more planets with periods less than 20 days, rising to 37$\%$ with periods less than 50 days.
This is comparable to 6$\%$ and 15$\%$ for 20 and 50 days for systems formed through planetesimal accretion.
These values show that in both pebble and planetesimal accretion scenarios, multi planet systems with orbits close to the central star, similar to \trap, can be considered common around low-mass stars.
However, it appears that systems formed through pebble accretion are slightly more compact, especially when looking at systems with a high number of planets.
This is understandable, since in the systems formed through planetesimals, there are extra dynamical interactions than in systems formed through pebble accretion.
The increase in dynamics arises from interactions with large numbers of planetesimals, which can act to break resonant chains of planets and lead to an increase in the number of collisions, reducing the overall planet number \citep{ColemanNelson16}.
Since chains of a larger number of planets will be more susceptible to an increase in dynamical interactions, it would be expected that these would be most affected, as can be seen from the statistics noted above.

\begin{figure}
\centering
\includegraphics[scale=0.6]{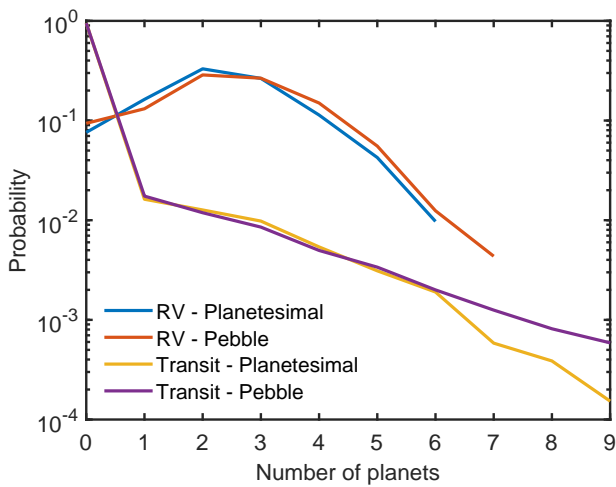}
\caption{The Probability of observing $N$ planets with periods up to 50 days in a simulated system, formed through either pebble or planetesimal accretion.
We show the probabilities for both the transit method, assuming that only planets with masses, $m_{\rm p} > 0.1\me$ may be observed, and the RV method, assuming a detection limit of 1 m/s.}
\label{fig:system_probs}
\end{figure}

Where Fig. \ref{fig:system_fraction} showed the number of planets in systems formed through either pebble or planetesimal accretion, they do not address the observability of these systems, so are not directly testable against observations.
Figure \ref{fig:system_probs} shows the probability of observing systems, either through the transit or radial velocity (RV) methods, containing $N$ planets that all have orbital periods less that 50 days.
For the RV method we assume that planets that induce a radial velocity in the star of K $\geq 1$m/s are detectable, whilst for the transit method we assume that planets more massive than 0.1$\me$ have radii large enough to be observed.

When comparing the probabilities for detecting $N$ planets through the RV method to the transit method, we see that the curves in Fig. \ref{fig:system_probs} have distinctly different shapes.
For the RV method, some systems are not observable since no planets are massive enough to induce a strong enough RV signal, amounting to 7.6 and 9.4 $\%$ for the planetesimal and pebble accretion scenarios respectively.
However, for the transit method, the number of systems that are not observable rises to $95 \%$, with little difference between both scenarios.
This dramatic difference in the observability of planetary systems through either the RV or transit methods, is due to the inclination of a planet with respect to the line-of-sight.
Since the effects of inclination are much more extreme for the transit method, then it is unexpected that the probability of observing at least a single planet is less than that compared with the RV method.
Compound the effect of a single planet's inclination with the mutual inclinations between planets in a single system, then it is clear that the observability of a larger number of planets will decrease as the number of planets increases.
This can be clearly seen in the transit probabilities in Fig. \ref{fig:system_probs}, where for both the pebble and planetesimal accretion scenarios, the probability of observing $N$ planets decreases with increasing $N$.

Since the RV method is only weakly dependent on inclination (especially for low inclinations), single planet systems are not the most common.
In fact, two and three planet systems from our simulations are most likely to be observed using the RV method.
Figure \ref{fig:system_probs} shows that these systems of two or three planets are more likely to be seen using the RV method than for the transit method.
Even for systems with 5 or more planets, the probability of being able to observe them through the RV methods is larger than through the transit method.
This however, is only due to the detection limits we have imposed on the observability of such systems, and does not take into account the difficulties that arise when disentangling multiple signals.
The disentanglement of multiple signals is considerably more complicated for the RV method, as opposed to the transit method, especially when the planets involved are in mean-motion resonance \citep{Anglada-Escude2010}.
When considering these difficulties, we expect the probability of observing a system with a large number of planets through the RV method to be much reduced to that shown in Fig. \ref{fig:system_probs}.

When increasing the planet number further to systems of 8 or more planets, we see that the probability through the RV method drops to zero, since none of the systems formed contained this number of planets that were detectable with our detection limits (K $\geq 1$m/s).
Generally in systems with a large number of planets, a significant proportion of these planets will have masses in the range $0.1 \me \le \mpl \le 0.5 \me$, which unless these planets are extremely close to the central star, they will not induce a strong enough RV signal to be observed.
However, since the planetary systems formed tend to be coplanar, they can be observed through the transit method, as is shown at the bottom of Fig. \ref{fig:system_probs}.
This makes the transit method more likely to observe systems with a larger number of planets, e.g \trap.

Given the harsh probability of observing a planet with the transit method, Fig. \ref{fig:system_probs} shows that the RV method is much more likely to yield results when observing low-mass stars.
However in terms of the yield of the number of planets in a system, the transit method can be favourable, especially if the systems are coplanar or in resonance.
In fact, if you apply the RV detection limits to the \trap ~system (using masses from \cite{Grimm18}), then only three planets would be observed, consistent with the expectations in Fig. \ref{fig:system_probs}, yet 7 planets have been observed with the transit method.
It is also interesting to note that due to the apparent magnitudes of stars in the sky, there are many more targets through the transit method, since it can probe to fainter magnitude stars than the RV method, whilst attaining sufficient signal-to-noise ratios that yield results.
This would increase the yield of planets found through transit observations, compared to RV observations, that could counteract the large differences in detection probabilities shown in Fig. \ref{fig:system_probs}.

It is interesting to note that when comparing the probabilities in Fig. \ref{fig:system_probs} to the actual distributions of planets up to 50 days, shown in both panels of Fig. \ref{fig:system_fraction}, there is great disparity.
For example, the majority of systems that could be detected will contain only a small number of planets, one planet being most probable in the transit method, and two -- three planets in the RV method.
However, Fig. \ref{fig:system_fraction} shows that systems with four -- six planets with periods less than 50 days are most common in both scenarios.
This disparity between the observed systems and the actual systems makes it extremely difficult to determine the planetary distributions that forms the observations.
Figure \ref{fig:system_fraction} also shows that the distribution of planetary systems is complex.
Generally as the number of planets remaining in a system increases, the number of these systems decreases.
But, the rate of this decrease appears non-linear from our simulated systems.
For example, in the pebble accretion scenario, there is a slight decrease of only 4 $\%$ in the number of nine-planet systems compared to four-planet systems.
In the planetesimal accretion scenario however, the differences in the distributions is much larger, with an 18 $\%$ difference between nine-planet and six-planet systems.
The differences in the distributions and the disparity with the observed probabilities therefore makes it extremely difficult to obtain in the present day, the distribution of planetary systems.

\begin{figure*}
\centering
\includegraphics[scale=0.55]{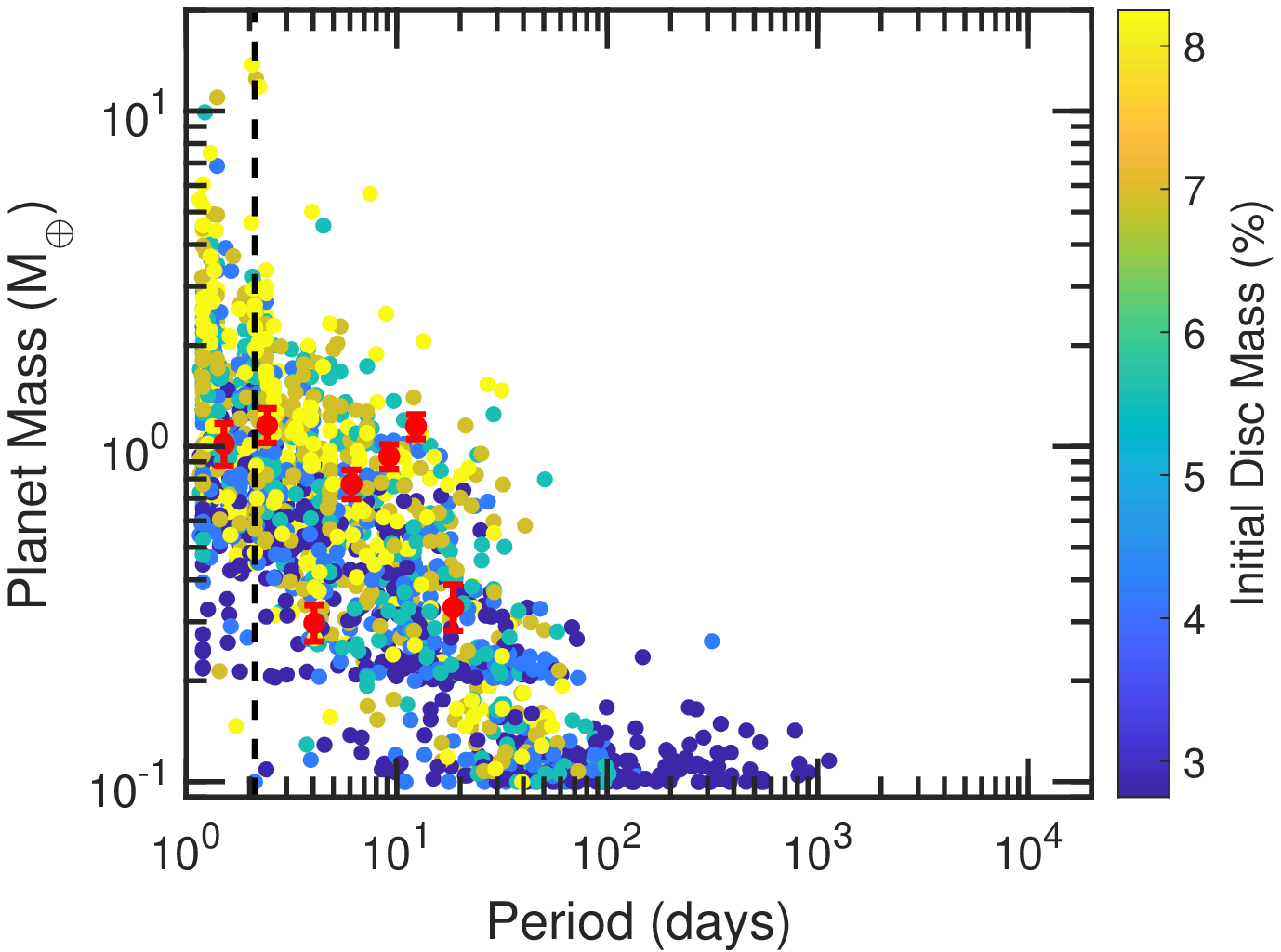}
\hspace{1cm}
\includegraphics[scale=0.55]{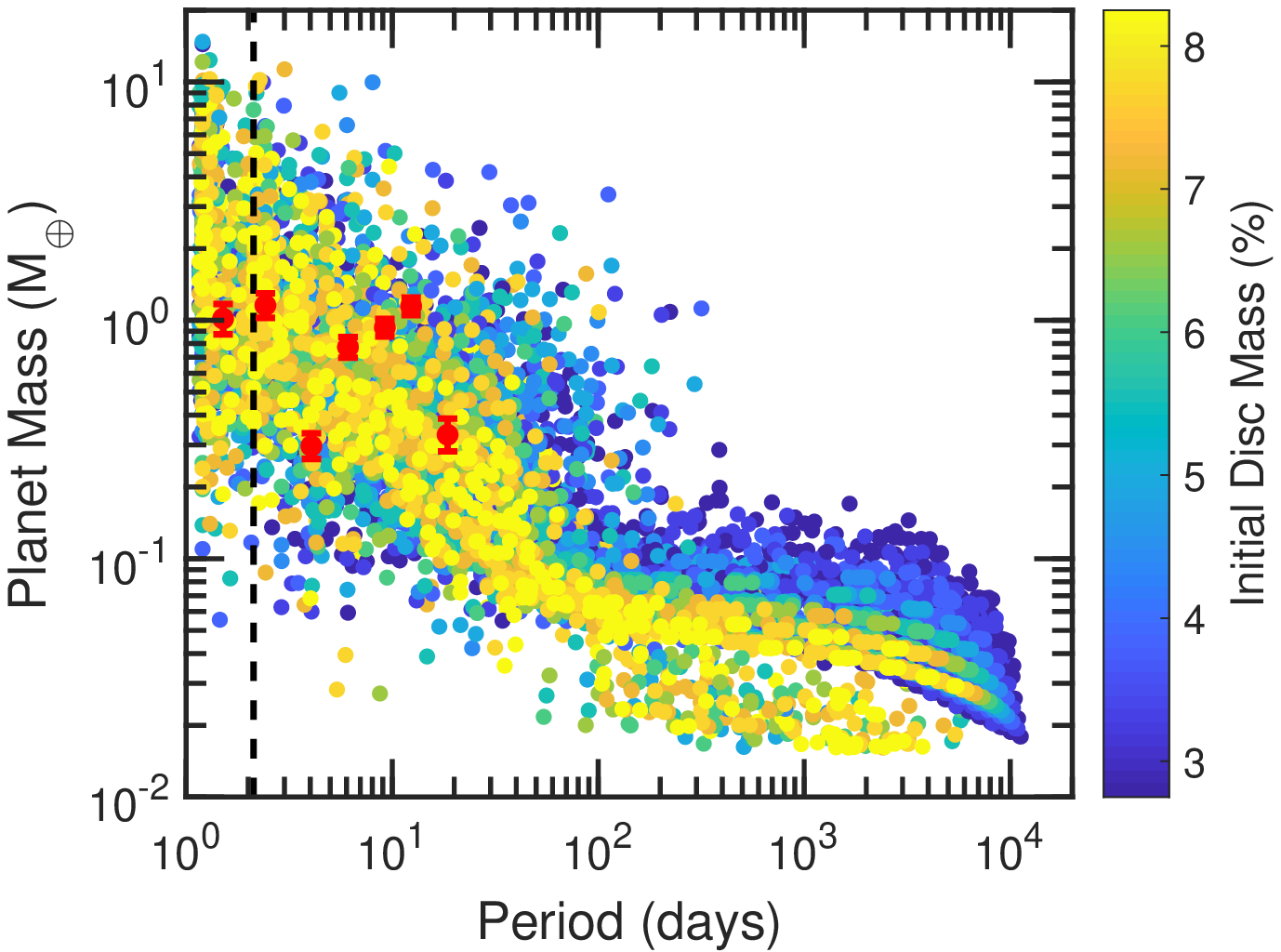}
\caption{A mass versus period diagram showing the the surviving simulated planets from the planetesimal accretion scenario ({\it left panel}), and the pebble accretion scenario ({\it right panel}).
Marker colour denotes the initial mass of the disc as a percentage of the stellar mass.
Red markers denote the masses and periods of the observed \trap ~planets with error bars \citep{Grimm18}.
The dashed line indicates the disc inner edge of 0.015$\au$ in the simulations.}
\label{fig:all_mvp}
\end{figure*}

In terms of the observability of systems formed through pebble or planetesimal accretion, there appears to be little difference.
With the RV method, the planetesimal accretion scenario gives a slightly higher probability for two planet systems, and a slightly lower probability for systems with four or more planets.
For the transit method, the probabilities for the two scenarios are similar for small numbers of observed planets.
However when comparing systems with larger numbers of planets, e.g. 7 or more, we see that the probability of observing these systems if they formed through pebble accretion is larger.
This is mainly due to these systems being more coplanar that their planetesimal accretion counterparts, since in the systems formed through planetesimal accretion, increased dynamical interactions with planetesimals will slightly increase the mutual inclinations between the planets, making systems with large numbers of planets less likely to be observable.
 
\subsubsection{Initial Disc Mass}

As well as looking at the properties of the planetary systems that form, it is also important to look at the planets individually.
In Fig. \ref{fig:all_mvp}, we show all of the surviving planets in a mass versus period plot for the planetesimal accretion scenario (left panel) and the pebble accretion scenario (right panel).
The colour of each planet denotes the initial disc mass, as a percentage of the central stellar mass.
We also show the \trap ~planets in red using using the periods and masses from \citet{Grimm18}, as well as their respective error bars.
As is clear from both panels of Fig. \ref{fig:all_mvp}, the \trap ~planets are embedded amongst the simulated planets, in terms of both mass and period.
Since there was insufficient solid material in the vicinity of the \trap ~planets in the simulations of either scenario, the simulated planets must have formed elsewhere and migrated in to their final locations.
Indeed, this is case, as the planets formed outside of the iceline, and then migrated in close to the central star in the form of resonant chains, similar to that discussed in Sects. \ref{sec:pltml_example} and \ref{sec:peb_example}.

It is also clear to see in both panels of Fig. \ref{fig:all_mvp}, that the more massive the planet, the closer to the star they typically orbit.
Since more massive planets migrate inwards faster, due to stronger Lindblad torques, this explains why they tend to be found closer to the star than less massive planets, as is evident by the super-Earth population of planets ($\mpl>3\me$) orbiting very close to their central star, with most of the population having periods < 10 days.
These more massive planets would have also forced other planets on interior orbits to migrate all the way on to the central star, leaving only the most massive planet behind, orbiting near the inner edge of the disc, along with less massive planets at longer periods.
When looking at the initial disc mass that formed these more massive planets, it is unsurprising to find that they formed in the most massive of discs examined.
This is unsurprising, since these discs have the largest abundance of solid material, either in the form of pebbles or planetesimals available for accretion by the planetary embryos.
Those super-Earths in the pebble accretion scenario (right panel) that are seen with larger periods typically became super-Earths after the end of the disc lifetime, where a resonant chain of planets became unstable, resulting in collisions between planets, forming the surviving super-Earths \citep{ColemanNelson16,Izidoro17,Lambrechts19}.

For less massive planets, both panels of Fig. \ref{fig:all_mvp} show that planets with masses $0.2\me\le\mpl\le2\me$ typically have orbital periods up to $\sim 100$ days.
For both scenarios, this appears to be independent of the initial disc mass, since in all discs simulated, once a planet reaches a mass $\mpl >0.5$--$1 \me$, its corotation torque saturates resulting in only inward migration.
In the pebble scenario, this one-way migration of the planets in this mass range is typically lower than the pebble isolation mass (eq. \ref{eq:peb_iso_mass}). This inward migration therefore slowed the accretion of pebbles, resulting in the planet masses being well below the pebble isolation mass. For the planetesimal accretion scenario, the planets underwent a similar experience, where here they migrated away from the planetesimal feeding zones before they were emptied. Even though the planets would be migrating into new feeding zones, their mass growth was still hindered, yielding similar results to the pebble accretion scenario. Therefore it seems that the migration tendencies of the planets, and accordingly the underlying disc properties, are the main sources of the distributions seen in fig. \ref{fig:all_mvp}.

Another effect is that whilst these planets are migrating inwards, they typically force other planets to migrate in with them in resonant chains, allowing the lower mass planets to migrate to periods they would otherwise have not been able to reach \citep{Alibert13}.
This migration scenario can explain the resonances in the \trap ~system as well as the differences in mass between \trap ~d and its neighbours, i.e. it was forced to migrate through interactions with its more massive outer companions.

For planets that have masses less than $0.2\me$, they underwent very little migration individually, since the torques that the disc could exert on these planets were very weak.
In both scenarios, these planets were only able to accrete limited amounts of solid material, more evident for the planetesimal accretion scenario in the left panel of Fig. \ref{fig:all_mvp}, where the planets were unable to even double their initial mass of 0.1 $\me$.
We find that this mode of limited accretion corresponds to the initial disc mass that is seen for these planets.
Typically, these planets were found in systems that had very small discs, where the supply of planetesimals was limited, and thus planetary growth was stemmed, similar to the `{\emph limited planetary growth}' scenarios observed in \citet{ColemanNelson14,ColemanNelson16}.
This is especially true for those planets of this mass range found with periods greater than 100 days.
However for those planets that appear with periods less than 100 days, and with high initial disc masses, their lack of mass growth has an alternative story.
In the planetesimal accretion scenario, even though these planets formed in more massive discs, their feeding zones of planetesimals, once emptied, were unable to be replenished.
This lack of replenishment arose from more massive planets accreting a significant number of planetesimals, whilst trapping others in mean motion resonances, preventing the feeding zones of the smaller planets from being replenished.
The other effect of these more massive planets, is as they migrated in towards the central star, they regularly trapped the smaller planets in resonance and forced them to migrate with them as part of a resonant chain.
This is the case for such planets in the pebble accretion scenario in the right panel of Fig. \ref{fig:all_mvp}.
Occasionally some of these planets would be scattered to larger periods, exterior to the resonant chains, where they could dampen their eccentricities and inclinations, without undergoing significant migration.

The scattering of low-mass planets is also seen in the pebble accretion scenario, where the planets at the bottom of the right panel of Fig. \ref{fig:all_mvp} are located.
Typically these planets formed closer to the central star, still outside the iceline, but were scattered out to larger orbital periods by more massive planets \citep{Fogg}.
This normally occurred after pebble accretion had ceased, i.e. the pebble growth front reached the outer edge of the disc, and as such, these planets were unable to further grow in mass.
It is also noted that these planets appear in the more massive discs.
This is understandable since the pebble surface densities are higher in more massive discs, allowing more massive planets to form and undergo significant migration towards the central star.
With more massive planets forming, the systems tend to be more dynamically unstable, which gives rise to further collisions and scattering events, forming this population of very low mass planets.

\begin{figure*}
\centering
\includegraphics[scale=0.5]{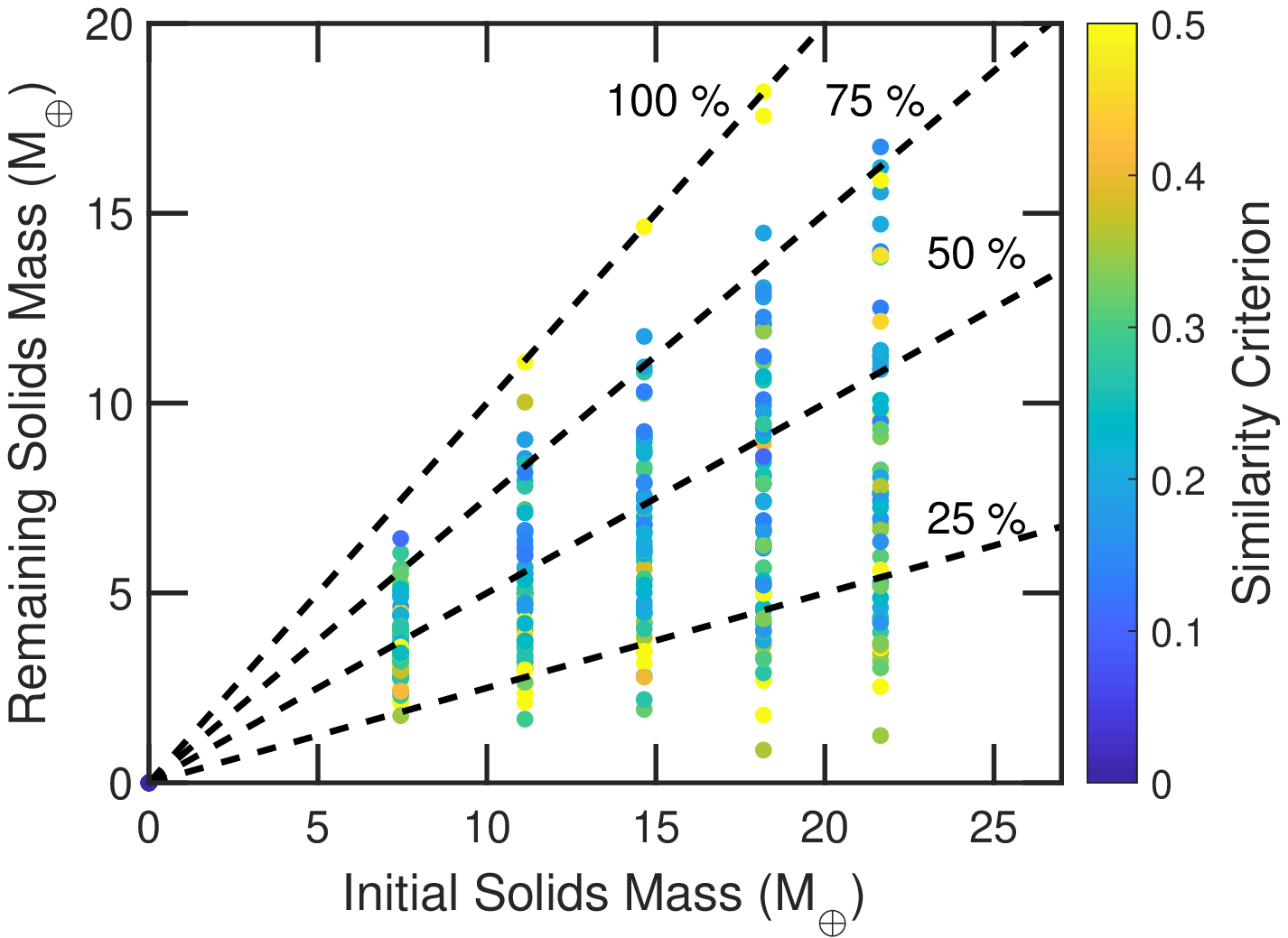}
\hspace{1cm}
\includegraphics[scale=0.5]{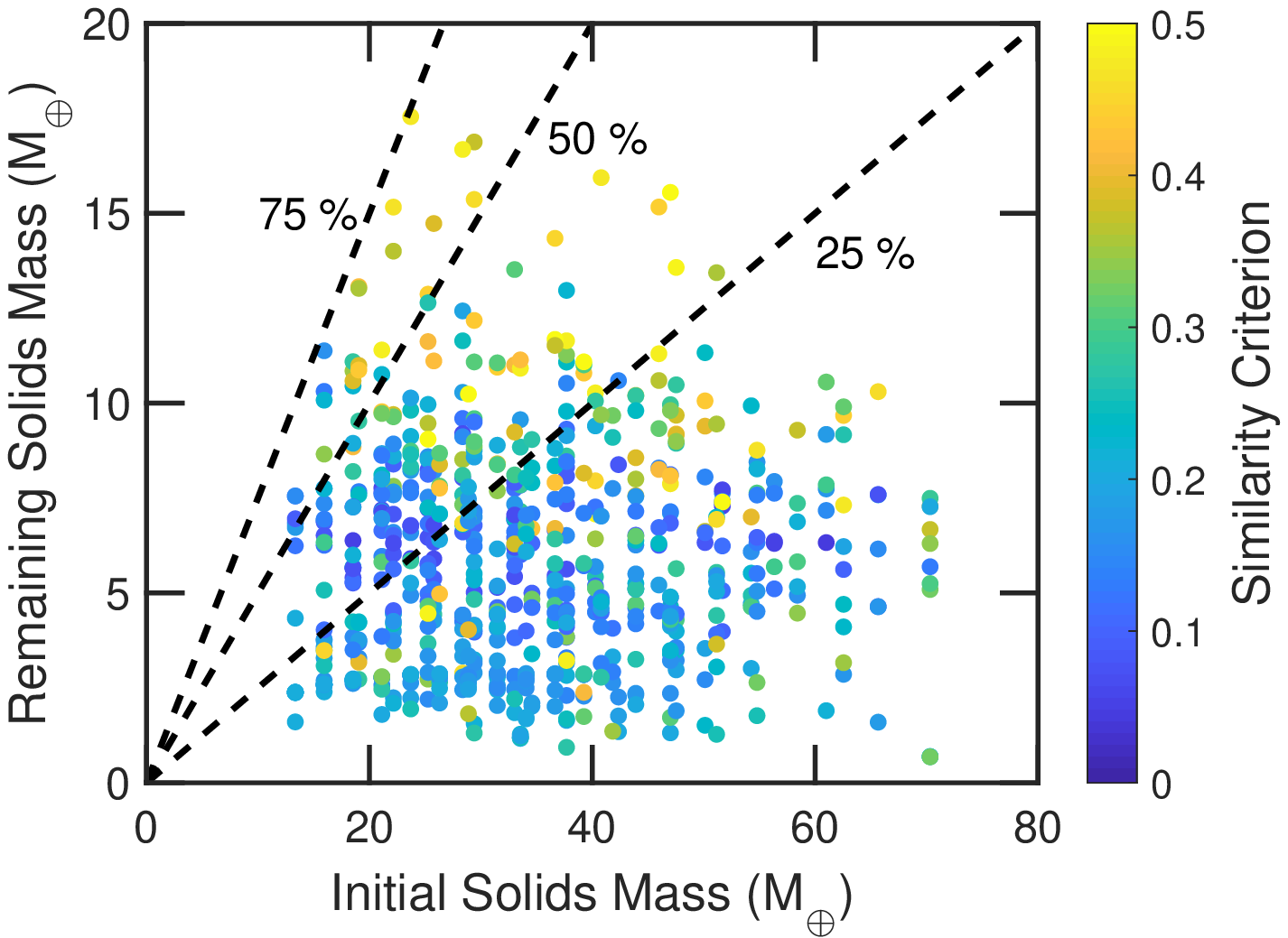}
\caption{The mass of solids remaining at the end of each simulation as a function of the initial solid mass.
Simulations from the planetesimal accretion scenario are shown in the {\it left panel} whilst those from the pebble accretion scenario are shown in the {\it right panel}.
All solids lost include those that have been ejected from the system and lost to the central star.
The dashed lines show the 25, 50, 75 and 100 $\%$ margins.
The colour coding in both panels corresponds to similarity criterion (eq. \ref{eq:distance}).}
\label{fig:efficiency}
\end{figure*}

\begin{figure*}
\centering
\includegraphics[scale=0.5]{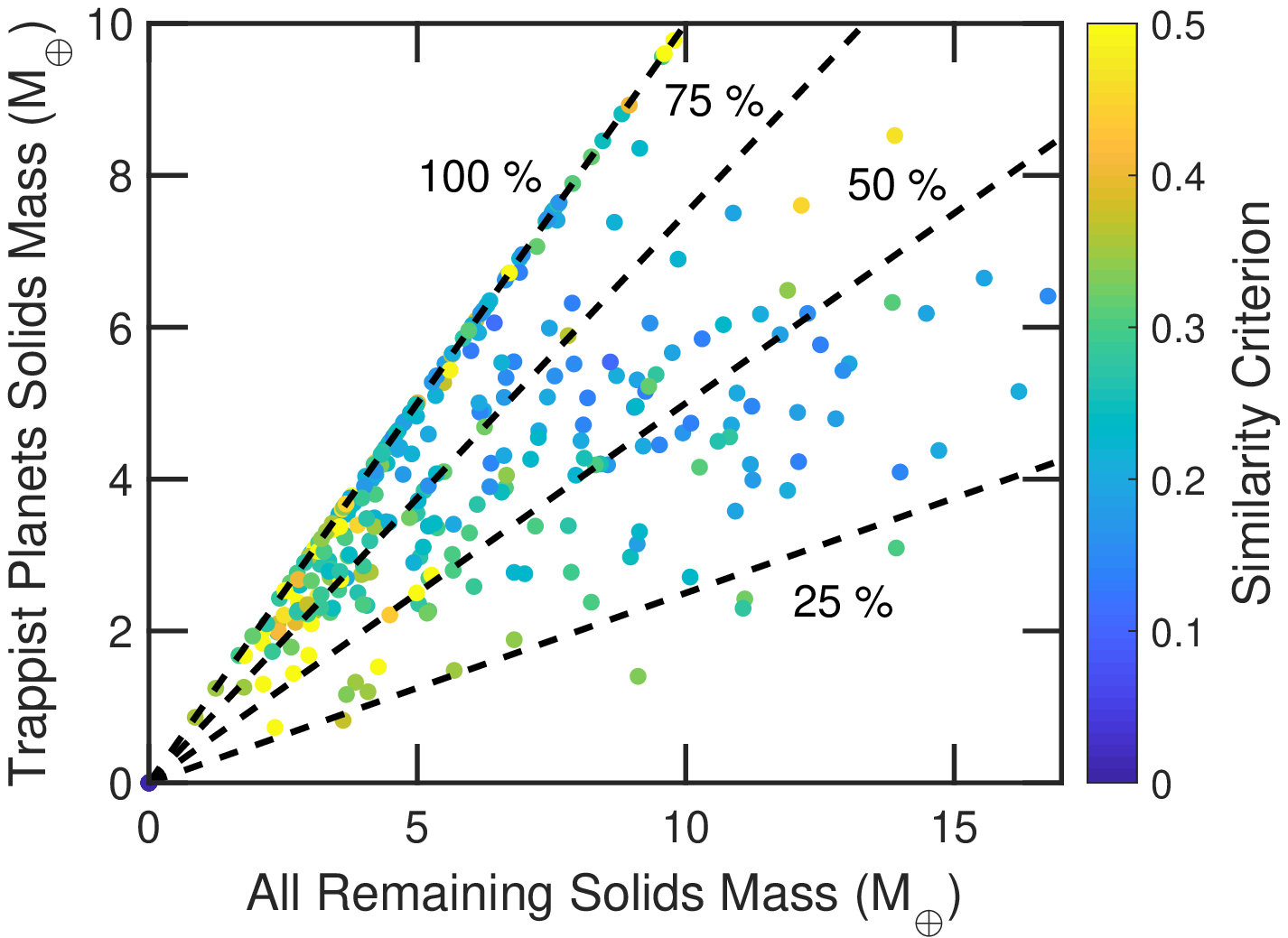}
\hspace{1cm}
\includegraphics[scale=0.5]{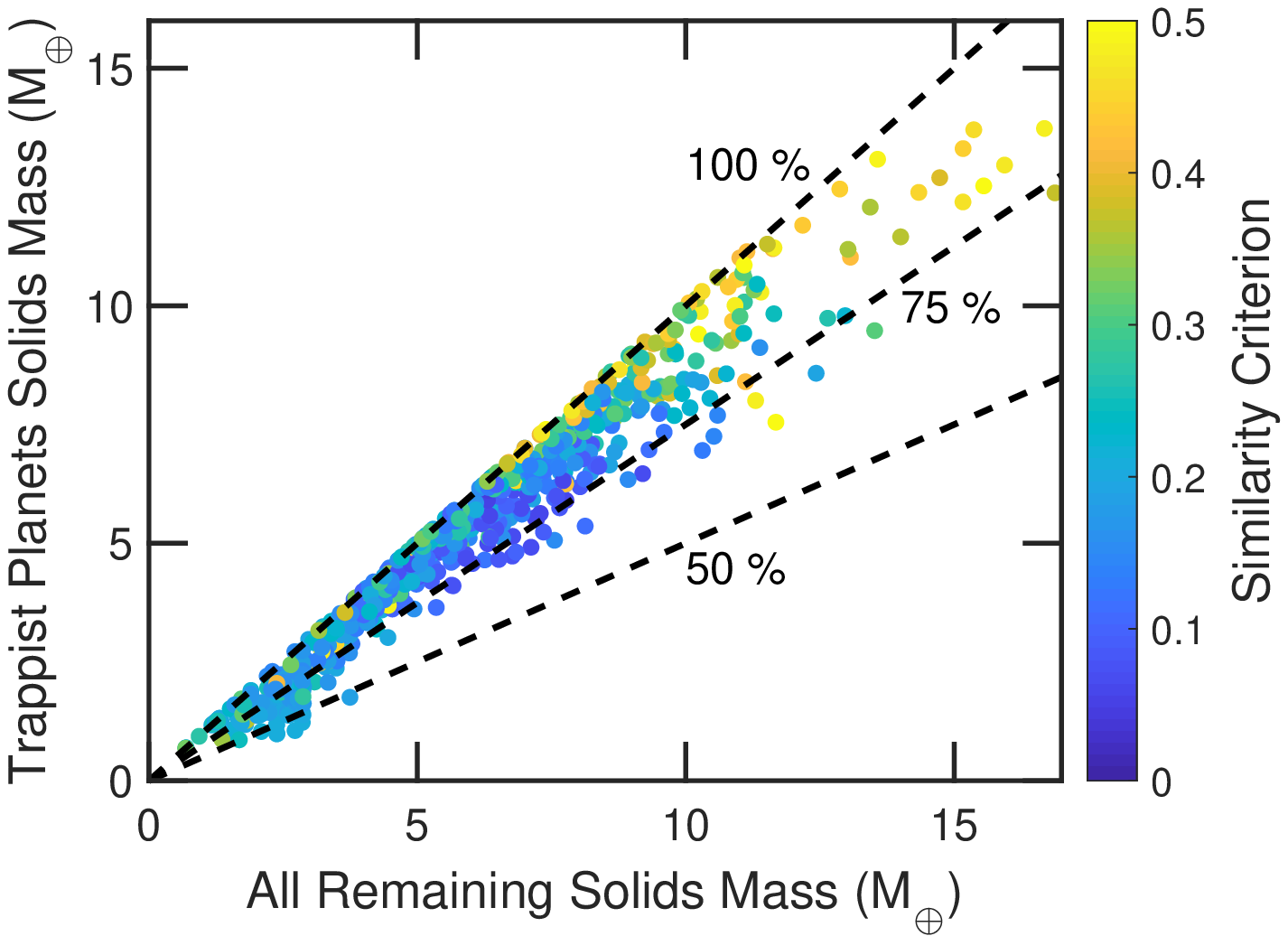}
\caption{The mass of remaining solids locked up in planets with orbital periods less than 20 days, against all remaining solid mass.
The {\it left panel} is for systems formed through planetesimal accretion, with the {\it right panel} being for systems formed through pebble accretion.
The dashed lines denote fractions of 25--100 $\%$, and the colour coding corresponds to the similarity criterion (eq. \ref{eq:distance}).}
\label{fig:trappist_percentage}
\end{figure*}

When comparing the pebble accretion scenario to the planetesimal accretion scenario, Fig. \ref{fig:all_mvp} shows that there are very little differences between the planets formed in the two scenarios.
Irrespective of the initial disc mass, the behaviours of the planets is similar.
However, when looking at the amount of solid material remaining versus what was initially in each scenario, there appears to be a significant difference.
Figure \ref{fig:efficiency} shows the fraction of solids remaining in the systems for the planetesimal accretion scenario (left panel), and the pebble accretion scenario (right panel).
The dashed lines represent masses where 25--100 $\%$ of initial solid material remains at the end of each simulation.
It is clear that for the planetesimal accretion scenario, generally the amount of solid material remaining is between 25--75 $\%$ of what was initially in the simulation, with an average of 45 $\%$.
This remaining material is locked up within planets making up the planetary systems and any remaining planetesimals.
The material lost from these systems includes any planets or planetesimals that have been impacted on to the central star, or ejected from the system.
Whilst the systems formed through planetesimal accretion retain a significant proportion of their initial solid material, those formed through pebble accretion generally retain less.
From the right panel of Fig. \ref{fig:efficiency} we see that the majority of systems only retain between 10--50 $\%$ of their original material, with an average of 21 $\%$ remaining.
Whilst the planetesimal accretion scenario assumed that 50 $\%$ of the amount of solid material was available for accretion, i.e. locked up in planets or planetesimals, the pebble accretion scenario allowed this value to vary between 50 and 90 $\%$, as this is shown to have an effect on final planets in the systems \citep{Brugger18}.
This increase in available material is why the initial amounts of solids in the pebble accretion scenario are much larger than the planetesimal accretion scenario.
Nevertheless, it appears that even with the extra material, the conversion of the pebbles into planets is small, and a lot of the pebbles simply drift through the system on to the central star.
This gives rise to the decrease in conversion efficiency that is seen in the right panel of Fig. \ref{fig:efficiency}.

Given that Fig. \ref{fig:efficiency} shows that there is substantially more solid material remaining in systems formed through planetesimal accretion than through pebble accretion, it is interesting to see where this leftover material is located within the system.
Figure \ref{fig:trappist_percentage} shows the mass within systems of all planets with periods less than 20 days versus the remaining solid mass in the entire system.
The planetesimal accretion scenario is shown in the left panel, and the pebble accretion scenario is shown in the right panel.
It is clear that for the pebble accretion scenario, except for a few systems, the planets with orbital periods less than 20 days encompass between 75 and 100 $\%$ of the total mass in their systems, with an average of 88 $\%$.
The remaining mass in these systems is locked up in low mass planets on longer periods as can be seen in the right panel of Fig. \ref{fig:all_mvp}.
This is unsurprising, since once planets gain sufficient mass, they begin to migrate away from their initial locations, heading inwards toward the central star.
The migration of these planets stalls near the inner edge of the disc, giving rise to the final planetary systems such as that discussed in Sect. \ref{sec:peb_example}.
In comparison, the systems formed through planetesimal accretion have a much greater spread of the final mass of objects throughout the system.
The left panel of Fig. \ref{fig:trappist_percentage} shows that in the majority of systems, 50--100 $\%$ of the mass is locked up in objects with periods less than 20 days, with an average of 75 $\%$.
The mass in these systems that is not located within a period of 20 days, is typically made up of low-mass planets such as those seen at the bottom of the left panel of Fig. \ref{fig:all_mvp}, and of planetesimals scattered throughout the system.
Over time these planetesimals would collide and erode, forming a debris disc, that can be observed.
Since pebble discs should erode on short time-scales, their debris disc signatures should be short-lived, so if debris discs were to be observed around low-mass stars, this could indicate that planetesimal accretion may be preferable to pebble accretion around low-mass stars.
Interestingly, dust belts have been tentatively observed with ALMA around Proxima \citep{Anglada17}, though their exact nature has yet to be determined.

\subsubsection{Water Fraction}

\begin{figure*}
\centering
\includegraphics[scale=0.55]{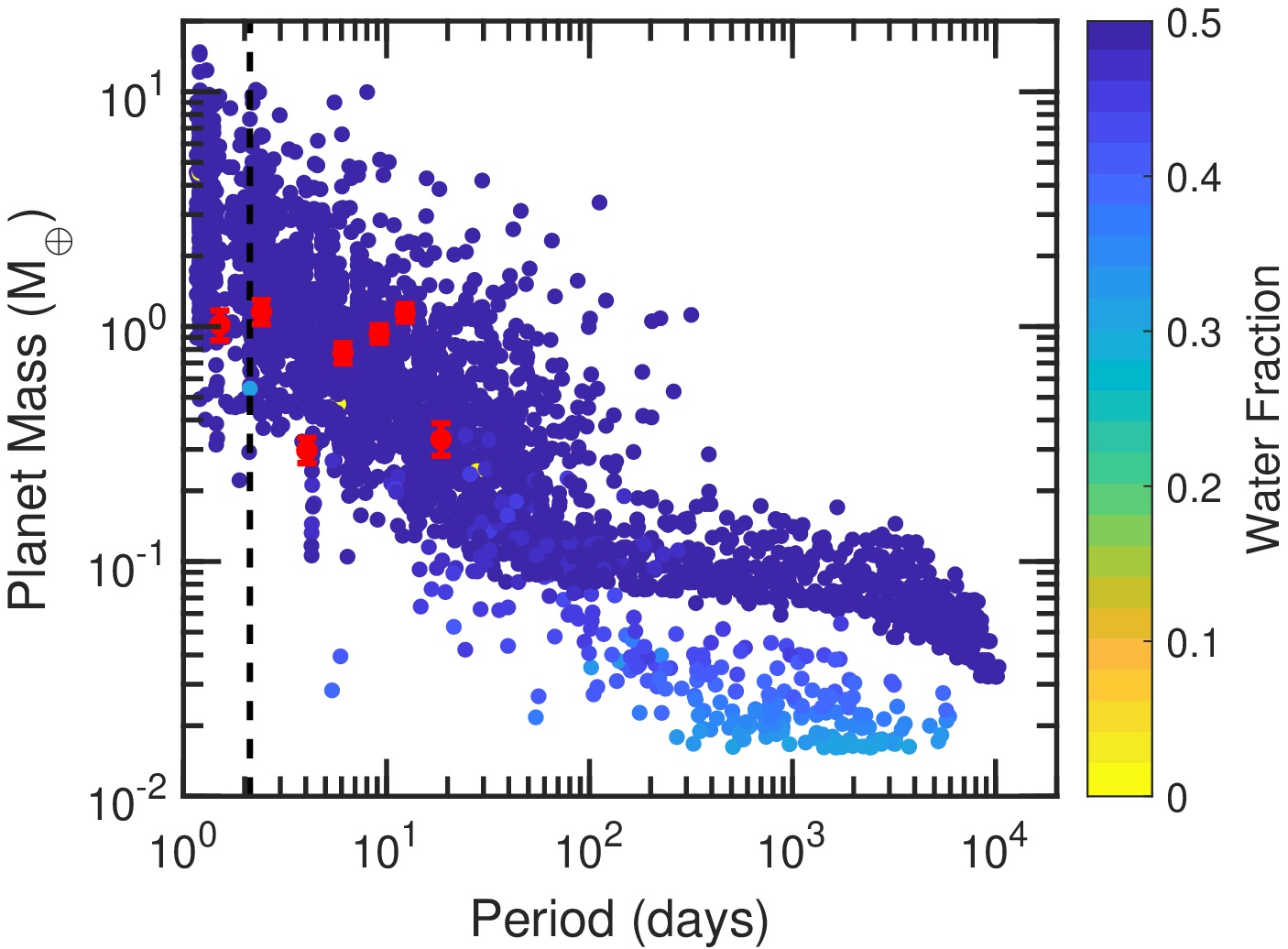}
\hspace{1cm}
\includegraphics[scale=0.55]{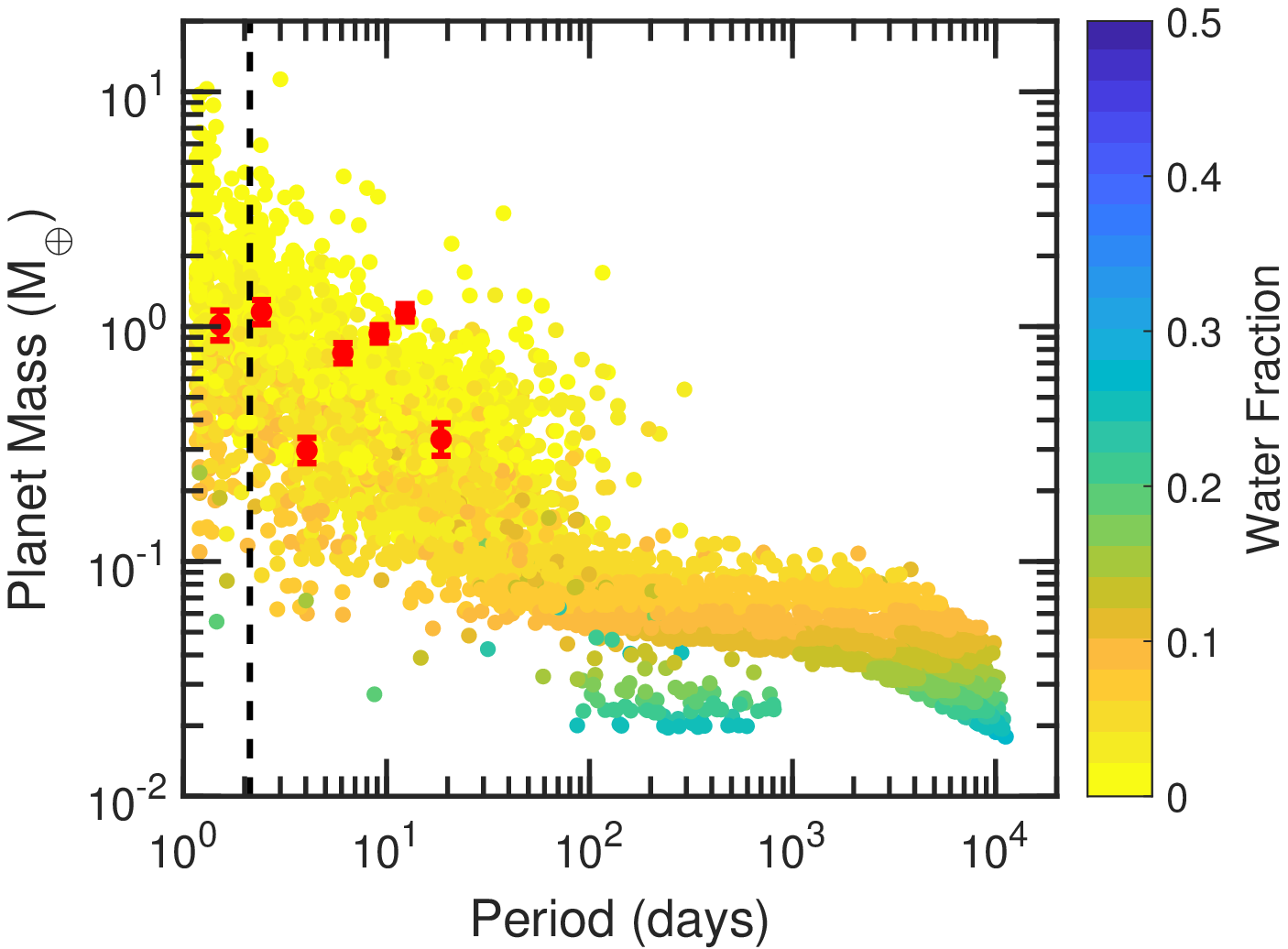}
\caption{Mass versus period plots showing the the surviving simulated planets from the pebble accretion scenario.
Marker colour denotes the final water fraction of the surviving planets.
Red markers denote the masses and periods of the observed \trap ~planets with error bars \citep{Grimm18}.
The \emph{left panel} shows the mass versus period diagram for simulations where the potential effects of pebble ablation and their efficient recycling back into the disc is not included.
The \emph{right panel} includes the potential effects of pebble ablation and the efficient recycling of the envelope back into the disc (see Sect. \ref{sec:ablation}).}
\label{fig:peb_water}
\end{figure*}

As well as comparing the planet masses and periods, it is also important to examine their compositions.
Where a planet accretes its solid material will have a considerable influence on its composition, since the material that is accreted will be reminiscent of the local disc material.
Since the disc has varying temperatures and pressures, the composition of the solid material accreted will vary throughout the disc \citep{Thiabaud14}.
In our simulations we focus on the amount of water ice that is accreted by the planets.
If the planet accretes solid material inside the water iceline, then there will be negligible amounts of water ice accreted since it will have sublimated.
Conversely, if the planet accretes solid material outside the water iceline, then a significant fraction would be comprised of water ice, depending on what the assumed ice-to-rock fraction is \citep{Lodders2003}.
The composition of a planet can also heavily influence its long term evolution, since the planets undergo photoevaporation from the central star over Gyr time-scales \citep{Owen17,Jin18}.
Depending on the strength of the photoevaporation, and the initial amount of water ice available for photoevaporation, planets can evolve and lose mass at different rates, also affecting their dynamical evolution within the planetary system.

Therefore, in Fig. \ref{fig:peb_water} we show the same mass versus period plot as the right panel of Fig. \ref{fig:all_mvp}, but instead the marker colour now denotes the final water fraction of the planets formed through the pebble accretion scenario.
Blue colours show planets that are water rich, whilst yellow show those that are water poor.
The left panel of Fig. \ref{fig:peb_water} shows the simulations that don't include the effects of ablation in the planet's envelope, with the right panel containing the ablative effects (see Sect. \ref{sec:ablation}).
Since most of the planets in the left panel of Fig. \ref{fig:peb_water} accreted the majority of their solid material outside the iceline, they are shown as being water rich.
The exception to these planets, being those at the lower region of the left panel of Fig. \ref{fig:peb_water}, where the planets have lower water fractions.
This is due to these planets accreting a significant amount of their material inside the iceline, where the pebbles are water poor.
Since the pebble surface densities inside the iceline are small, these planets remain low in mass.
Interactions with more massive planets as they migrate through the disc, scatters a number of these low mass planets to larger orbital periods, where they remain due to weak migration forces.

The right panel of Fig. \ref{fig:peb_water} shows the water fraction of planets formed in the simulations when ablation is taken into account.
The effects of ablation are obvious here when comparing the right panel to the left panel of Fig. \ref{fig:peb_water}.
Where the effects of ablation are neglected (left panel), the water fraction of the planets is typically around $\sim 50\%$, that being the assumed composition of pebbles outside the iceline \citep{Lodders2003}.
However, in the right panel, most of the planets with masses greater than Mars (0.1 $\me$) have water fractions less than $5\%$.
Even though the planets accreted pebbles outside the iceline, most of the water content in the pebbles ablated in the planets' atmosphere.
This water, now within the planets' atmosphere was then recycled back into the disc \citep{Ormel15b,Cimerman17}.
The remaining solid component of the pebble, assumed to be rocky is then accreted on to planetary cores, allowing them to grow whilst simultaneously reducing their water fraction.
This ablative effect gives rise to the low water fractions seen in the planets in the right panel of Fig. \ref{fig:peb_water}, with water fraction decreasing as the mass increases.

Whilst the planets that formed here, when neglecting the possible effects of ablation and envelope recycling, attained water fractions of $\sim 50\%$, the planets formed in \citet{Schoonenberg19} contained significantly reduced water fractions.
In their simulations the initial water rich planetesimals ($50 \%$ water) formed just exterior to the ice line, before accreting pebbles from their local vicinity.
As these planetesimals migrated inwards (only inwards migration was allowed), they accreted initially wet pebbles, and then once the planets crossed the ice line, dry pebbles, before eventually ending the simulations on orbits close to the central star.
Since the planetesimals accreted both wet and dry pebbles, this enabled them to reduce their water fraction, where depending on model parameters, the planets comprised of water fractions between 7 and 39$\%$ with their fiducial model being 10 $\%$.
These values are significantly lower than the results shown here, and is mainly due to the types of pebbles they accrete.
For the planets in the simulations here, some are initially placed interior to or surrounding the ice line, compatible with the initial positions in \citep{Schoonenberg19}.
However since the models presented here include outwards migration, these planets begin to migrate outwards once they accrete a sufficient mass of pebbles, until they reach a zero migration region\footnote{It is also noted that planets that formed exterior to the ice line would also migrate inwards to these locations before having their migration stalled. This effect was also not included in the models of \cite{Schoonenberg19} and would affect the types of pebbles that are accreted by the planets.}, typically located at transitions in opacity \cite{ColemanNelson14}.
This outwards migration is caused by the local viscosity and entropy gradients strengthening the planets corotation torques, making them stronger than their Lindblad torques.
This results in those planets mainly accreting wet pebbles, and as remaining water rich, $\sim 50\%$.
Given that it only seems that the main reason for the discrepancies in the water fractions between the models presented here and those of \cite{Schoonenberg19} arise from the types of pebbles that are accreted and the inclusion of outward migration, it would be fair to suggest that should the models of \cite{Schoonenberg19} include outwards migration, or should the models here neglect it, then the water fractions between the planets that form near the ice line would be similar.

Whilst the planets formed through pebble accretion contained either water poor or water rich planets, depending on whether ablation of pebbles was included in the envelope, planets formed through planetesimal accretion are found to be water rich.
In fact the planets are found to have water fractions around $\sim50\%$, since these planets accreted the majority of their solid material outside the iceline.
Interior to the iceline, there was insignificant amounts of planetesimals available for accretion initially, and any planetesimals that were interior to the iceline would migrate through gas drag forces onto the central or star, or be forced to migrate as exterior planets migrated inward towards the star, shepherding the planetesimals along with them.
As a consequence, there is little opportunity for planets to reduce their water content \citep{AlibertBenz17}.
Even when accounting for the ablation of the planetesimals in the planets' envelope, which we neglect for planetesimals here, the water fraction would be little changed.
This is due to the planets' envelope not being massive enough to thermally ablate the planetesimals, nor massive enough to destroy the planetesimals through mechanical destruction \citep{Alibert17}.
Therefore this accretion scenario leads to the very high water fractions, similar to the left panel of Fig. \ref{fig:peb_water}, and would suggest that if the \trap planets formed through planetesimal accretion, then they should be extremely wet.
However, as the planets evolve over time, they absorb radiation from their central star that can act to remove some of the water \citep{Bolmont17}.
This will lead the planets to having smaller observed water fractions than they had formed with, possibly blurring out their formation pathway in the process.

When comparing the water fractions of the surviving planets, we therefore find that planets formed through planetesimal accretion tend to be more water abundant than those formed through pebble accretion.
This is only when comparing the planets that formed through planetesimal accretion to those that formed through pebble accretion but included the effects of ablation as the pebbles passed through the planet's envelope.
If the effects of ablation of pebbles are neglected then the water fractions of the planets formed through pebble accretion are comparable to those that formed through planetesimal.
Since we expect pebbles to undergo ablation, the right panel of Fig. \ref{fig:peb_water} should be more representative of the water fraction of planets formed through pebble accretion, and as such should make for a better comparison with planets formed through planetesimal accretion.
Therefore, if the water fraction of the \trap ~planets or other planets around low-mass stars, e.g. Proxima b, could be determined, then this could hint at the mode at which planets accrete their solid material, i.e. through either pebble accretion or planetesimal accretion.
Obviously, long-term evolution processes can also affect the observed water fraction, and blur out the differences between the two accretion processes, i.e. FUV photolysis \citep{Bolmont17}, as could the treatment of ablated material within the planet's gaseous envelope.
For our models, the impact of ablation is dependent on the assumption that the solid material that is ablated is efficiently recycled back into the protoplanetary disc \citep{Ormel15b,Cimerman17,Bethune19,Bethune19b}.
If the ablated material is not efficiently recycled back into the disc \citep{Lambrechts17,Kurokawa18}, then it would contribute to the heavy elements in the envelope \citep{Lambrechts14b,Venturini15}, and could also settle towards the centre of the planet, affecting the planet's water content.

\begin{figure*}
\centering
\includegraphics[scale=0.55]{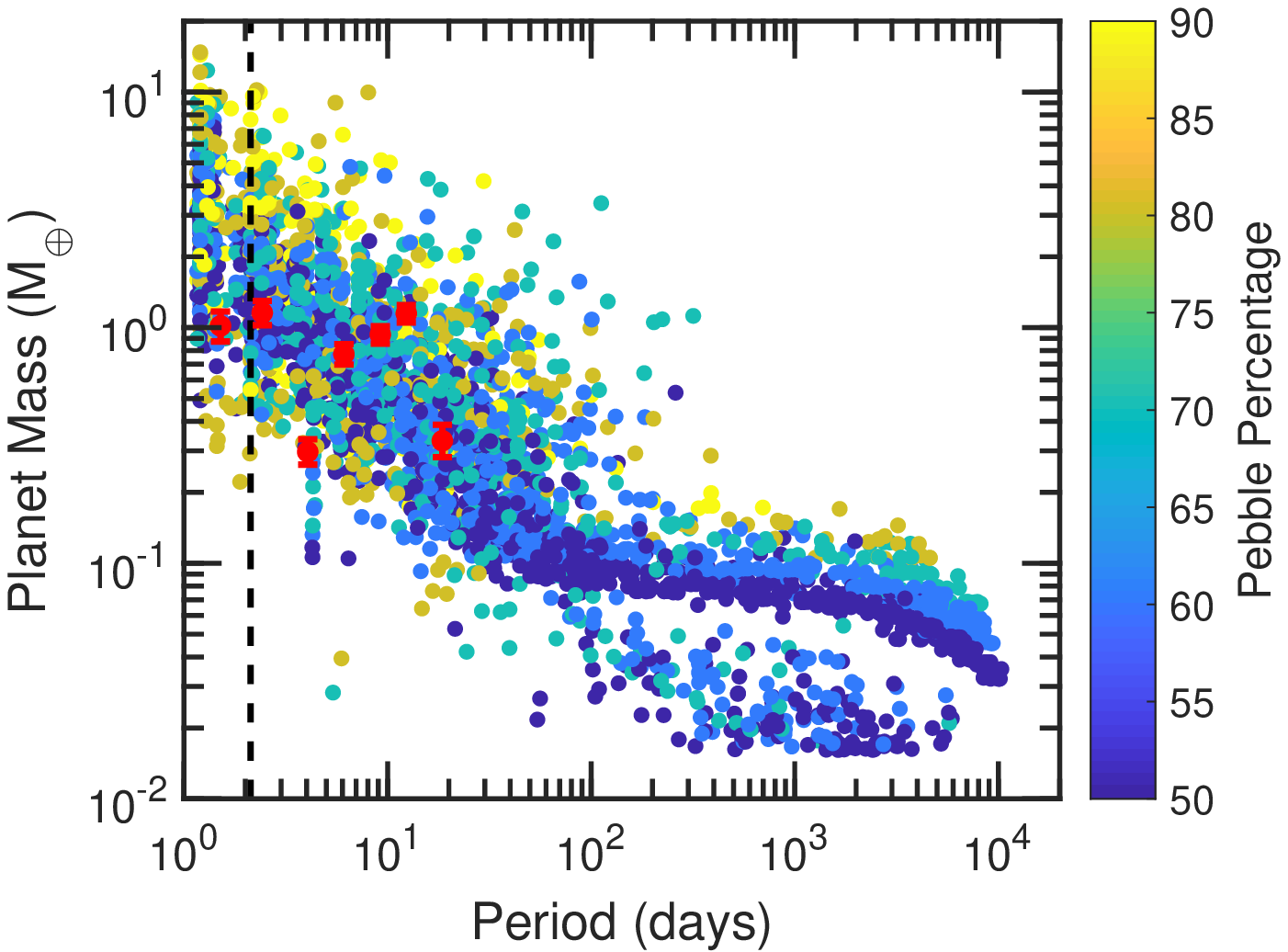}
\hspace{1cm}
\includegraphics[scale=0.55]{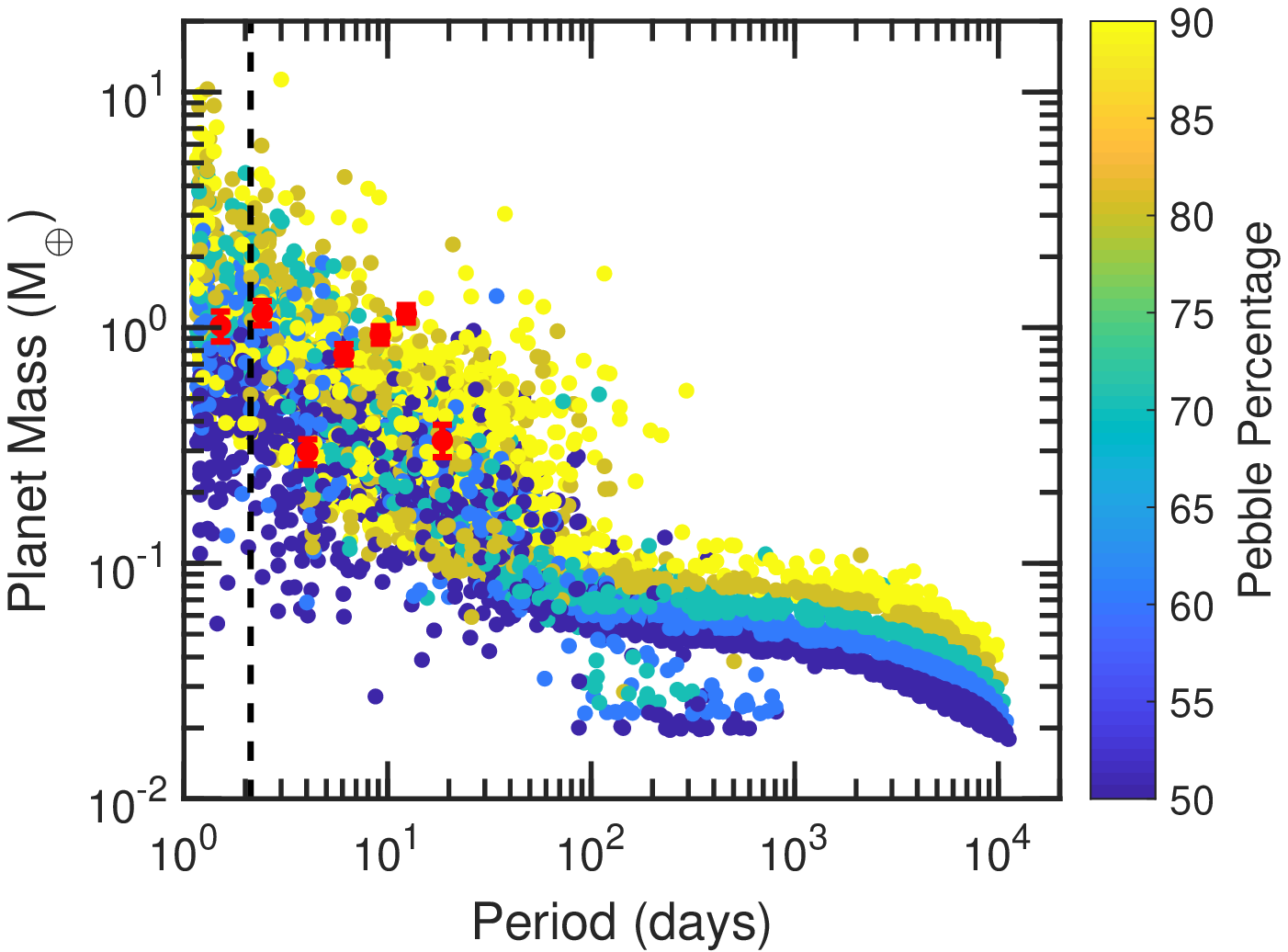}
\caption{Mass versus period plots showing the the surviving simulated planets from the pebble accretion scenario.
Marker colour denotes the percentage of solids that is comprised of pebbles.
Red markers denote the masses and periods of the observed \trap ~planets with error bars \citep{Grimm18}.
The \emph{left panel} shows the mass versus period diagram for simulations not including the effects of pebble ablation.
The \emph{right panel} includes the effects of pebble ablation.}
\label{fig:peb_perc}
\end{figure*}

\subsubsection{Pebble Percentage and Planetesimal size}

Figure \ref{fig:peb_perc} shows the same mass versus period plot as in Fig. \ref{fig:peb_water}, but the marker colours here denote the percentage of solids in the disc that is composed of pebbles, with the remaining percentage being that of dust.
The left panel of Fig. \ref{fig:peb_perc} shows the simulated planets not including the effects of ablation in the planet's gaseous envelope, with the right panel showing the planets including the effects of ablation.
As the percentage of solids comprised in pebbles increases, the supply of pebbles increases, allowing planets to attain higher masses \citep{Brugger18}.
This can be seen in the planets in both panels of Fig. \ref{fig:peb_perc}, more pronounced for the lower mass planets to the right of both panels that underwent little migration or dynamical interactions.
The increase in mass as the pebble percentage increases is particularly noticeable in the right panel of Fig. \ref{fig:peb_perc} since the accretion rate of pebbles was reduced due to ablation in the planets gaseous envelopes, yielding lower mass planets that experienced very few dynamical interactions.
In regards to more massive planets, the effect remains but is less pronounced.
This is due to the increase in dynamical interactions between $\sim$Earth-mass planets than for Mars-mass planets.
These interactions induce more collisions which can blur the impact of different pebble percentages.

\begin{figure}
\centering
\includegraphics[scale=0.6]{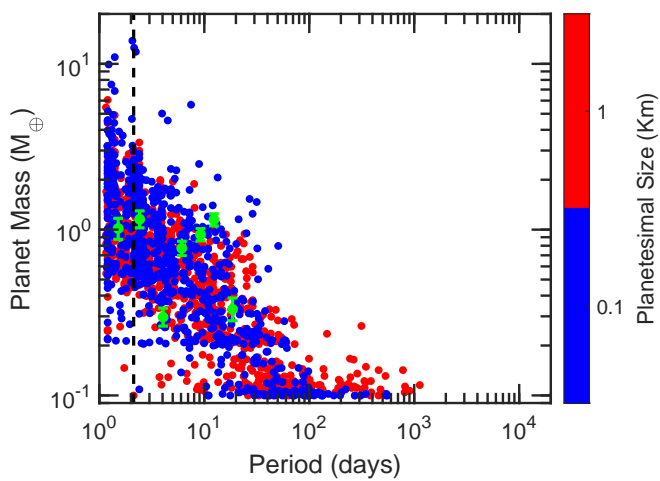}
\caption{Mass versus period plot showing the the surviving simulated planets from the planetesimal accretion scenario.
Marker colour denotes the planetesimal size in that simulation.
Green markers denote the masses and periods of the observed \trap ~planets with error bars \citep{Grimm18}.}
\label{fig:pltml_all_mvp_pltml}
\end{figure}

For the planetesimal accretion scenario it is interesting to look at the effect of the planetesimal size, since this can dramatically affect the accretion rate of solids \citep{ColemanNelson16,ColemanNelson16b}.
Surprisingly, we find that the initial size of planetesimals seemed to have very little effect on the types of planetary systems that formed.
It is seen that systems that contained smaller, 100m planetesimals, did contain slightly more massive planets, and comprised more growth.
However in terms of the mass range, $0.2\me<\mpl<2\me$, the systems that contained 1km planetesimals match very closely those that contained 100 m planetesimals.
This can be seen in the mass versus period plot in Fig. \ref{fig:pltml_all_mvp_pltml}, where the colour marker shows the planetesimal size that formed those particular planets.
The blue and red dots overlap in the mass range mentioned above, and there is an abundance of blue points, denoting 100 m planetesimals, at high masses close to the central star, and conversely an abundance of red points, denoting 1 km planetesimals, at low masses covering periods up to 1000 days.
The increase in planetary mass growth as planetesimal size decreases is consistent with previous works \citep{ColemanNelson16,ColemanNelson16b}, but these simulations show that the sensitivity on the planetesimal size is not as pronounced for low-mass stars as they are for Solar-type stars.
The reason for the lack of sensitivity to the planetesimal size, is due to the initial solid surface density slopes (see Sect. \ref{sec:pltml_setup} for explanation of the surface density slope).
With steeper slopes, e.g. -3, there is a greater concentration of planetesimals near the iceline, where planets tend to migrate to.
With this higher concentration, the accretion rates are higher, allowing the planets to become more massive, which leads to them more quickly dominating the planetesimal dynamics, through scattering, rather than by gas drag.
When looking at the planet masses in the simulations with flatter initial profiles, e.g. -1.5, the difference in masses between 100m and 1km simulations is more pronounced, and more consistent with previous works \citep{ColemanNelson16,ColemanNelson16b}.

Whilst in this work we chose planetesimal sizes of 100m or 1km, it has been shown that the streaming instability mainly forms planetesimals of sizes between 10km and 150km \citep{Johansen15,Simon16,Li19}.
Using planetesimals of larger sizes has been found to reduce the accretion efficiency onto planets, where if planetesimals of 10 km were used, very little planetary growth was seen  \citep{ColemanNelson14,ColemanNelson16}.
This was mainly due to the weak damping from gas drag on those planetesimals, that allowed them to retain significant eccentricities which increases their relative velocities compared to planetary embryos, thus reducing accretion efficiencies.
Smaller planetesimals experience more gas drag, and thus the accretion efficiencies onto planetary embryos is increased.
Therefore for planetesimal accretion scenarios to utilise the enhanced accretion efficiencies of smaller planetesimals, some mechanism has to be assumed that converts the larger planetesimals into their smaller counterparts.
Collisional stirring of planetesimals, especially if they form in dense narrow rings, will convert some of the larger planetesimals into smaller ones.
Obviously this conversion will occur on some time-scale, of which we assume to be similar to that required for the formation of the $\sim$Mars mass embryos that we initiate our simulations with.

\subsubsection{Mean Motion Resonances}
\label{sec:comp_resonance}

\begin{figure}
\centering
\includegraphics[scale=0.65]{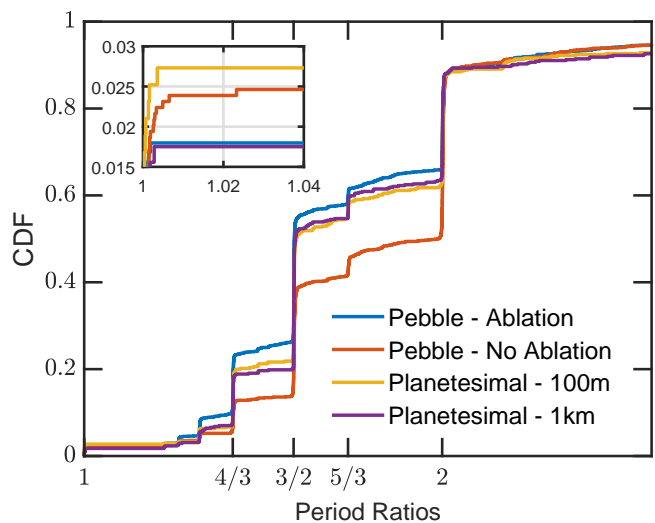}
\caption{Cumulative distribution functions of period ratios between neighbouring planets with periods less than 20 days.
Blue and red lines show the period ratios for pebble accretion simulations that include/not include the effects of ablation.
The yellow and purple lines show the period ratios for the planetesimal accretion scenario for the planetesimals of size 100m and 1km respectively.
The inset plot zooms in on the coorbital region of the distributions.}
\label{fig:ratio_cdf}
\end{figure}

The planets in \trap ~are all found to be in mean-motion resonance.
This is normally a common outcome of planet formation scenarios that include migration \citep{ColemanNelson14,ColemanNelson16}.
As planets migrate, they can form resonant chains of planets that end up migrating in a resonant convoy \citep{Hellary}, typically with the most massive planet dominating the migration tendencies.
These resonant chains remain stable during the disc lifetime due to gas damping of eccentricities and inclinations and can remain stable after the disc has fully dispersed.
Typically though, the resonant chains found in planet formation scenarios, are first order MMRs, i.e. 2:1, 3:2, etc \citep{ColemanNelson16}.
However the \trap ~planets, though exhibiting these first order MMRs, also contains planets appearing to be in second and third order MMRs, 5:3 and 8:5 respectively.

When we examine the resonant structures coming from the pebble and planetesimal accretion simulation results, we find an abundance of first order MMRs, particularly 3:2 and 2:1, as expected from previous planet formation scenarios \citep{ColemanNelson16}.
Figure \ref{fig:ratio_cdf} shows cumulative distribution functions of the period ratios of all neighbouring planets with periods less than 20 days from all of the simulations.
The pebble accretion scenarios are shown in blue (with ablation) and in red (without ablation), whilst the planetesimal accretion scenarios are shown in yellow (100m planetesimals) and purple (1km planetesimals).
Each vertical jump shows where there is an abundance of planets with that specific period ratio with their interior neighbour.
This normally occurs at/near mean motion resonances, typically of first order, e.g. 3:2 or 2:1, but can occur away from the mean motion resonance locations, i.e. the observed abundance of neighbouring planets with period ratio $\sim2.2$ found in the {\it Kepler} data \citep{Steffen15}.

It can be seen in Fig. \ref{fig:ratio_cdf} that the first order resonances of 4:3, 3:2 and 2:1 are abundant with planet pairs in all of the populations.
These resonant pairs are normally in resonant chains of planets, undergoing smooth migration from the outer regions of the disc into periods close to the star.
For both planetesimal populations, and for the pebble population that included the effects of ablation, there appears to be very little difference between the period ratios of planet pairs.
This is somewhat unsurprising since the planets in each case have similar masses, and the resonant chains have similar properties.
If the mass of the planets was more sensitive to the planetesimal size used or the mode of accretion used, then there would be a greater disparity between the three distribution functions, since the stability of the system would decrease as the mass of planets, in identical resonances, increases \citep{PuWu2015}. 
This disparity can be seen when comparing the three distributions mentioned to the pebble population that doesn't include the effects of ablation.
For that distribution, the planet masses are slightly larger, since all of the pebble mass entering the planets envelope is accreted by the core.
With this increase in planet mass, the systems are generally more dynamically unstable, leading to an increase in the number of mutual interactions and collisions between planets, reducing the compactness of the system, that is seen in Fig. \ref{fig:ratio_cdf}.

Also seen in the resonant pairs in Fig. \ref{fig:final_systems} and in the four distribution functions in Fig. \ref{fig:ratio_cdf}, is the abundance of planet pairs occupying the second-order 5:3 resonance.
Typically, when planets migrate in resonant chains, they do so in first-order resonances, since their eccentricities are kept at small values through damping from the protoplanetary disc, which allows the planets to typically bypass second-order MMRs.
When studying the evolution of the planet pairs that are found in second-order MMRs, this is still the case.
Instead of forming the second-order resonances within a resonant chain, the planets here follow a different pathway.
Initially the planets are part of a larger resonant chain of planets, made up of first-order MMR.
As a chain goes unstable, interactions and collisions between planets occurs, possibly forcing planets out of resonance.
If this happens with a sufficient gas disc remaining, then the planets will migrate, and since they have non-negligible eccentricities, they are much more easily able to become trapped in second-order resonances, since the libration width of second-order resonances increases with eccentricity.
This is the case for all of the planet pairs here, and could give a possible hint as to the evolution of \trap-c and \trap-d that appear to be in a 5:3 resonance.
In terms of the occurrence of second-order resonances, Table \ref{table:MMR} shows that 7.3$\%$ and 9$\%$ of all mean motion resonances in the pebble and planetesimal accretion scenarios respectively are second order.
When only looking at planets close to the central star (periods < 20 d) and those with masses $\mpl > 0.1\me$, these rates rise to 9$\%$ and 9.8 $\%$ respectively.
Table \ref{table:MMR} also shows that 24$\%$ of pebble accretion systems contain at least one pair of planets in second order resonance, compared to 16$\%$ in the planetesimal accretion systems.
These rates do not change when only considering planets within 20 d, showing that most of the second order resonances are close to the star.
Also interestingly, in both scenarios, $\sim 95 \%$ of second-order resonances are part of a larger chain, again highlighting the formation process of these resonant configurations.

Relating to \trap, we see that planets `b' and `c' are seen to be close to the 8:5 resonance.
Third order resonances are extremely weak and very hard to attain in planet formation scenarios, since planets will typically migrate through the resonance before becoming trapped in the first order 3:2 resonance.
However, in the simulations we do find some planet pairs near the 8:5 resonance.
On closer inspection of these planet pairs, we find that instead of being in 8:5 resonance, both planets are in first order resonance with a third planet, closer in towards the star.
These resonances being 5:4 and 2:1 respectively.
Instead of the outermost planet migrating past the 8:5 resonance with the middle planet, it becomes trapped in the first order 2:1 resonance with the innermost planet, and now migrates as part of a chain.
It is only that the middle planet is in a 5:4 resonance with the inner planet, that the 8:5 period ratio between the middle and outermost planet appears, see for example `Pebble System 2' in Fig. \ref{fig:final_systems}.
This could have interesting consequences for the innermost \trap ~planets, as to whether there is another unseen planet, with a high inclination, or whether there was another planet at the formation of the system, but has since disappeared as the system has evolved over time, possibly due to tidal or magnetic migration \citep{Strugarek17}.

\begin{table}
\centering
\caption{Occurrence rate of second order and co-orbital mean motion resonances.
For the `With limits' case, we limit our sample to planets with periods less than 20 d, and masses greater than 0.1$\me$.}
\begin{tabular}{l c c c c}
\hline
 & \multicolumn{2}{c}{Pebbles} & \multicolumn{2}{c}{Planetesimals}\\
 & All & With limits & All & With limits \\
 \hline

\multicolumn{5}{l}{{\bf With respect to all MMR}}\\
Second Order &  $7.3\%$& $9.0\%$& $9.0\%$ &  $9.8\%$ \\
Co-orbital &  $3.1\%$& $4.0\%$& $2.2\%$ &  $2.6\%$\\
\multicolumn{5}{l}{{\bf At least one per system}}\\
Second Order &  $24\%$ & $16\%$ &  $23\%$ & $16\%$ \\
Co-orbital &  $10\%$ & $8.7\%$&  $5.9\%$ & $5.2\%$ \\
\multicolumn{5}{l}{{\bf Part of a resonant chain}}\\
Second Order &  $95\%$ & $95\%$& $96\%$ & $95\%$\\
Co-orbital &  $62\%$& $48\%$& $82\%$ & $71\%$\\
\hline        
\end{tabular}
\label{table:MMR}
\end{table}

It is also interesting to note that we find a number of co-orbitals in both the pebble and planetesimal accretion scenarios, with approximately $2.2\%$ and $2.6\%$ of the planet pairs with periods less than 20 days.
When looking throughout the entire system, these rates rise to $3.1\%$ and $4\%$ respectively.
Since our initial conditions do not allow the in-situ formation of co-orbital bodies near the Lagrangian points of existing planets \citep{Laughlin02,Beauge07,Lyra09}, and with the existence of a chaotic area surrounding the co-orbital domain preventing their capture by slow migration \citep{Wisdom80}, they must have formed through a different mechanism.
We find in the simulations that these co-orbital pairs typically formed where there were numerous dynamical instabilities allowing planets orbits to cross, before strong eccentricity damping allowed them to settle into either trojan or horseshoe orbits \citep{CresswellNelson2006,cressnels}.
Interestingly, the stabilising effects of resonant chains on co-orbital configurations as discussed in \cite{Leleu19} can also aid in the formation of the co-orbital resonances, and we find that in both scenarios, most of the co-orbital pairs within 20 d are in resonant chains (82$\%$ and 71$\%$ for pebble and planetesimal systems respectively).
Interestingly these rates drop when removing the 20 d limit, to 62$\%$ and 48$\%$ respectively, showing that there are more isolated co-orbital pairs, where these pairs are found to form in a resonant chain, but were left behind as the rest of the chain migrated in towards the star.
We note however that in the co-orbital case, the local perturbations of the protoplanetary disc caused by both planets affects the stability of the co-orbital resonance that cannot be properly modelled by a 1D disc model \citep{Cresswell09,Pierens14,Broz18,Leleu19}.
Results on this configuration are hence mainly given in order to compare the models to one another.
The occurrence rates are shown in Table \ref{table:MMR}

Whilst the planets in the \trap ~system have not been shown to be in two-body resonance, they have been shown to be in three-body resonance \citep{LugerTrappist1-h}.
We find in both sets of simulations that resonant chains are abundant, consequently resulting in numerous three-body resonances, see figs. \ref{fig:pltml_3body} and \ref{fig:peb_3body} for example.
Hence, it is unsurprising that the \trap ~planets are also found to be in three-body resonance.
These resonant chains can be modified over time through tidal evolution, which can affect their stability and two-body appearance \citep{Strugarek17,PapaloizouTrappist}.

\section{Conclusions and Discussions}
\label{sec:conclusions}
In this paper we have explored the formation of planetary systems around low mass stars, similar to \trap.
To form these systems we ran numerous N-body simulations using the Mercury-6 symplectic integrator that has been coupled with a self-consistent 1D viscously evolving disc model, and contains prescriptions for stellar irradiation, type I migration and the final dispersal of the disc through photoevaporation.
To form the planets in the systems we examined two different accretion scenarios: pebble accretion and planetesimal accretion.
In the planetesimal accretion scenario, we embedded planetary embryos amongst $\sim$2000 planetesimals, whilst in the pebble accretion scenario we included a pebble surface density and accretion model following \citet{Lambrechts14}.

In both scenarios, we find that systems similar to \trap ~are able to form in a diverse range of initial conditions.
This includes the initial disc mass, the size of planetesimals and the percentage of solids converted to pebbles.
In all simulations, these planets typically accrete the majority of their material outside of the iceline before migrating into the inner regions of the disc, to finish with periods comparable to those of \trap.
As the planets migrate, they form resonant chains and migrate in convoy with the most massive planet typically dominating the migration tendencies.
These resonant convoys remain stable after the end of the disc lifetime, since the planets are separated by a sufficient number of mutual hill radii \citep{PuWu2015}.

When comparing the Trappist-like systems from both scenarios with \trap, we find good agreement with planet masses, periods, and resonances with neighbouring planets.
In terms of resonances within each individual systems, we find that first-order resonances dominate, especially 4:3 and 3:2, but we also find second-order resonances appear fairly frequently.
These second order resonances form when a resonant chain goes unstable during the disc lifetime, and the remaining planets are able to slowly migrate into the second-order resonance, if they have significant eccentricity.
The only discernible difference between the two scenarios was the final water fraction of the planets.
Systems that formed through planetesimal accretion were water rich since most of the solid material was accreted outside the iceline.
The Trappist-like systems from the pebble accretion scenario were found to be water poor.
This was due to the water content ablating in the planet's envelope and then being recycled back into the protoplanetary disc.
This disparity in the water fractions was the only observable difference between the two scenarios in regards to \trap.

\subsection{Comparing Pebbles with Planetesimals}

When comparing the two scenarios against each other, we find that for most parameters the two populations compare favourably.
In terms of the average masses, eccentricities, inclinations, period ratios, and their respective dispersions, there was little to separate the systems formed by pebble accretion from those by planetesimal accretion with our chosen initial conditions.
One aspect where there was a small difference was in the gradient of planet mass as a function of period.
The systems formed through pebble accretion tended to have a slightly more negative gradient than their planetesimal counterparts.
This resulted in the pebble accretion systems generally having more massive planets close to the central star, and ever lower mass planets further out.
This could be an observable signature, however a large number of planetary systems would have to be observed to a detailed level, so as to gain adequate statistics.

The other main difference between the two scenarios was the number of planets orbiting within 50 days.
Whilst there were only small differences in the number of planets observed in systems from each scenario, the systems formed through pebble accretion tended to have slightly more planets than the systems formed through planetesimal accretion.
This increase in the number of planets is attributed to the reduction in dynamics in the pebble scenario, since in the planetesimal scenario, swarms of planetesimals can dynamically stir up resonant chains of planets, making them go unstable and cause collisions, and therefore reducing the number of planets.

When looking at the observability of the simulated systems, we find little difference between the two scenarios using both the RV and transit methods.
Using the transit method, there is only a $5 \%$ chance of observing a system with at least one planet from either scenario.
We find that one planet systems will be the most commonly observed, with systems with increasing number of planets less probable to be observed.
The slight increase in the number of planets in a system between pebble and planetesimal accretion scenarios can be seen here, but only when observing systems of seven or more planets.
Here systems formed through pebble accretion have a slightly higher chance of being observable.
Whilst the transit method finds a majority of one planet systems, the RV method finds that two or three planet systems are most likely.
This is mainly due to mutual inclinations between planets, that even though they are not high, are significant enough to affect the observability of the planets for the transit method.
On the flip side, if a system of a large number of planets is coplanar, then they could all be observed, e.g. \trap.
For the RV method we find that systems of six or more planets are very difficult to observe with current limits, since the planets are too small to induce an observable signature on the star.
To summarise both methods inadvertently miss planets when observing, mainly due to observational biases, and whilst it is more probable for the RV method to observe a system, generally these systems will be low in the number of planets, the transit method will have a much higher chance of observing plentiful planetary systems.

We also find that in the planetesimal accretion scenario, there is an abundance of leftover material on long period orbits, that could constitute a debris disc.
These debris discs are not seen in the pebble accretion scenario, since the majority of the large dust would have drifted through the system as pebbles and would either have been accreted by the planets, or the central star.
With the tentative debris disc signatures around Proxima \citep{Anglada17}, if more debris discs were found around low mass stars, this could point to planetesimal accretion as being a preferred mechanism for planet formation.

\subsection{Application to Other Observed Planetary Systems}
\begin{figure}
\centering
\includegraphics[scale=0.6]{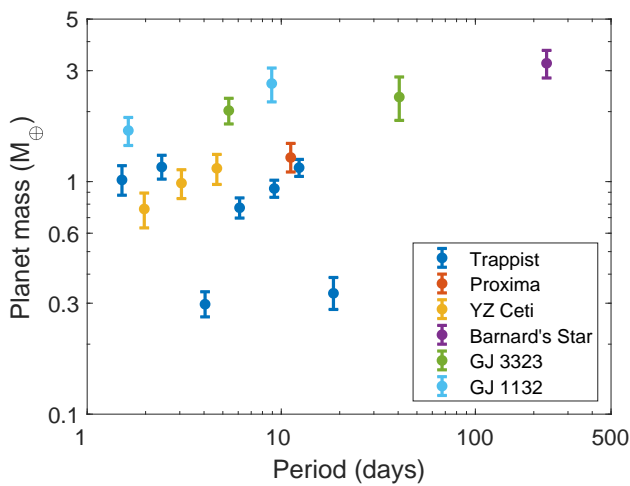}
\caption{A mass versus period plot showing planetary systems similar to \trap. Observational data was taken from exoplanet.eu.}
\label{fig:other_systems}
\end{figure}

Whilst this paper is focused on comparing the pebble accretion scenario to the planetesimal accretion scenario in the context of \trap, it is also interesting to briefly compare the simulation results to other planetary systems observed around low-mass stars.
Figure \ref{fig:other_systems} shows the masses and periods of other observed planetary systems, as well as that of \trap ~for reference.
For the low mass stars in the figure (Proxima and YZ Ceti), it is clear to see that there is considerable overlap in the masses and periods of the planets with those of \trap.
Proxima b \citep{Anglada2016} has very similar orbital properties to \trap-g, whilst the planets orbiting YZ Ceti \citep{Astudillo-Defru17}, have similar masses, periods and resonances to those seen in the \trap ~system, however the exact status of the YZ Ceti planets is still to be finalised \citep{Robertson18}.
These systems also look very similar to simulated systems from Sects. \ref{sec:planetesimals} and \ref{sec:pebbles}, especially that of YZ Ceti, but the fact that only one planet has been observed around Proxima is puzzling.
Simulated planetary systems show that we would expect there to be more planets in the Proxima system, most likely of similar or lower mass to Proxima b, however none have been observed.
This could be due to low mass of any expected planets, and to the disentangling the RV signals from planets in resonant chains.

When looking at stars of slightly more mass, Barnard's Star, GJ 3323, and GJ 1132, it is clear to see that these planets are equally more massive than the planets that form around low mass stars such as \trap.
Even though the increase in stellar mass is only a factor of 2 (GJ 1132 has a mass of 0.164 $\msun$), it could be expected that the planetary systems could look similar to \trap.
Indeed, the planetary system of GJ 1132 \citep{Berta-Thompson15,Bonfils18_GJ1132} does contain multiple planets orbiting within 10 days, but they are slightly more massive, and don't appear to be in resonance, much unlike the systems formed in the simulations presented here.
There are also two planets around GJ 3323 \citep{Astudillo-Defru17_GJ3323}, but they have orbital periods between 5--40 days, and masses of around 2 $\me$, meaning the system is slightly extended both in period and mass, when comparing to \trap.
More recently a super-Earth has been observed around Barnard's star \citep{Ribas18}, however unlike the planetary systems around GJ 1132 and GJ 3323, this planet has a period of 232 days, making it completely incompatible with the simulated planetary systems formed here.
Given the increase in masses and orbital periods of the planets around Barnard's Star, GJ 1132 and GJ 3323, and that these systems all seem to be metal-poor compared to \trap, it would seem that the planet formation processes that formed these planets could be slightly different, even though the change in stellar mass is modest.

\subsection{Overall Remarks and Future Work}

Whilst the disc model of \cite{ColemanProxima17} that we use here has an outer boundary of 10$\au$, it is possible that the the discs can be much larger in size.
This could possibly have implications on the amount of solids available for accretion, and how long a pebble growth front can continue to form pebbles that migrate into the inner system.
With these concerns in mind, we checked what effects of having a larger disc, out to $40\au$, but containing the same initial mass, would have on the pebble growth front and on planetesimal formation.
Having a larger disc would allow more time for planets to accrete pebbles, since the pebble growth front would have a further distance to traverse before reaching the outer edge of the disc.
However, given that the disc masses would be the same, the amount of pebbles produced would be significantly reduced, meaning fewer pebbles would be travelling past a planet.
When accounting for fewer pebbles travelling past a planet, but for a longer time, and calculating the total amount of pebbles that will flow past a planet throughout the pebble accretion regime, we find that the total amounts are the same irrespective of the disc size, as long as the disc masses are equal.
This leads us to conclude that even with larger protoplanetary discs, the pebble accretion scenario would yield similar to results to those in Sect. \ref{sec:pebbles}, with the only difference being the accretion rate onto the planets being subdued and occurring over a slightly longer time-scale.

As for the planetesimal accretion scenario, a more extended disc will have the effect of spreading the mass in planetesimals over a larger area, thus reducing their local surface densities.
This spreading of the planetesimals was partially explored here, where different power laws were used to spread the planetesimals, finding that the steeper slopes, i.e. higher concentrations of planetesimals, allowed for increased growth for the less mobile 1 km planetesimals.
With shallower slopes, planetesimal accretion rates decreased as a function of the planetesimal size and overall density as expected from previous works \citep{ColemanNelson16}.
Given that a more extended disc would reduced the local number of planetesimals, either a greater number of initial planetesimals, or a steep slope, e.g. -3, would be required to form planetary systems similar to \trap.
But given that the location that planetesimals form is still not well understood, steeper slopes in the planetesimal surface densities in the inner regions of the protoplanetary disc could be common \citep{Drazkowska17}.

Whilst this work was able to adequately form systems similar to \trap, it is by no means complete.
Further improvements to the models are required to not only improve their accuracy, but also their comparability to observations of the \trap ~planetary system.
In future work, we will aim to include the following improvements:\\
\indent(i) Since planetesimals are able to migrate inwards from beyond the iceline, in principle these should sublimate quite rapidly. We have not included submlimation in our models, but it is seen in the simulations that planets accreted icy planetesimals after they cross the iceline. In practice, some, if not all of the ice should have sublimated as it crossed the iceline, and therefore a model of planetesimal sublimation should be included for self-consistency.\\
\indent(ii) Include more up-to-date prescriptions for internal photoevapoation, as well as include appropriate prescriptions for external photoevporation to simulate high-energy cosmic rays entering the system from nearby high-mass stars.\\
\indent(iii) Incorporating a more realistic migration model that takes into account 3D effects \citep{Fung2015}, the influence of planet luminosity \citep{Benitez-LlambayMasset2015} and dynamical torques arising from the planet's migration \citep{Paardekooper2014,Pierens15,McNally18}.\\
\indent(iv) Include the effects of planet eccentricity and inclination on the accretion rate of pebbles from the protoplanetary disc \citep{Liu18,Ormel18}.\\
\indent(v) Include more accurate prescriptions for the recycling of planetary envelopes.

Additionally, Even though the systems formed here are numerous, they do not represent a full population synthesis of planetary systems around low mass stars, as we have not chosen initial conditions (e.g. disc mass, metallicity, lifetime) from observationally motivated distribution functions.
This will be the subject of future work, which should allow for a more complete comparison between the pebble and planetesimal accretion scenarios around low mass stars.
It should also allow for the determination of how rare the \trap ~planetary system is, and whether more systems like it should be observed.

\begin{acknowledgements}

The authors acknowledge support from the Swiss NCCR PlanetS and the Swiss National Science Foundation.
We would also like to thank the anonymous referee for their constructive comments that improved the quality of this paper.

\end{acknowledgements}

\bibliographystyle{aa}
\bibliography{references}{}

\end{document}